\documentclass[sigconf,screen]{acmart}
\usepackage{hyperref}
\usepackage{graphicx}
\usepackage{balance}
\usepackage{caption}
\usepackage{subcaption}
\usepackage{xcolor,colortbl}
\usepackage{fancybox}
\usepackage{adjustbox}
\usepackage{multirow}
\usepackage{rotating}
\usepackage{tikz,pifont}
\usepackage{amsfonts,amsmath}
\usepackage{hyperref}
\usepackage{xspace}
\usepackage{multirow}
\usepackage{makecell}
\usepackage{enumitem}
\usepackage{booktabs}

\newlist{rqs}{enumerate}{1}
\setlist[rqs,1]{label=\textbf{RQ\arabic*.},ref=\textbf{RQ\arabic*}}

\newcommand{\tool}{FairRF\xspace}
\newcommand{\rqoneshort}{Comparison with base strategies\xspace}
\newcommand{\rqone}{To what extent is \tool able to improve fairness and effectiveness in predictions compared to base approaches?\xspace}
\newcommand{\rqtwoshort}{Comparison with algorithms variations:\xspace}
\newcommand{\rqtwo}{How does \tool compare against variations of the MOEA algorithm that employ different base classifiers?\xspace}
\newcommand{\rqthreeshort}{Comparison with SOTA:\xspace}
\newcommand{\rqthree}{How does \tool compare against state-of-the-art bias mitigation methods for bias mitigation with single sensitive attributes?}
\newcommand{\rqfourshort}{Intersectional fairness:\xspace}
\newcommand{\rqfour}{How does \tool compare against state-of-the-art bias mitigation methods for intersectional bias mitigation?}
\newcommand{\replpackage}{\cite{replpackage}}

\newcommand{\isBetter}{\cellcolor[HTML]{AAD5D3}}
\newcommand{\isWorse}{\cellcolor[HTML]{FBC44D}}

\definecolor{lightblue}{HTML}{AAD5D3}
\definecolor{orange}{HTML}{FBC44D}

\AtBeginDocument{%
  \providecommand\BibTeX{{%
    \normalfont B\kern-0.5em{\scshape i\kern-0.25em b}\kern-0.8em\TeX}}}

\setcopyright{acmcopyright}
\copyrightyear{2026}
\acmYear{2026}
\acmDOI{XXXXXXXXXXX}

\acmConference[ICSE-SEIS '26]{2026 IEEE/ACM 48th International Conference on Software Engineering}{April 12--18, 2026}{Rio de Janeiro, Brazil}

\begin{document}


\title{\tool: Multi-Objective Search for Single and Intersectional Software Fairness}

\author{Giordano d'Alosio}
\orcid{0000-0001-7388-890X}
\affiliation{%
  \institution{University of L'Aquila}
  \city{L'Aquila}
  \country{Italy}
}
\email{giordano.daloisio@univaq.it}

\author{Max Hort}
\orcid{0000-0001-8684-5909}
\affiliation{%
  \institution{Simula Research Laboratory}
  \city{Oslo}
  \country{Norway}
}
\email{maxh@simula.no}

\author{Rebecca Moussa}
\orcid{0000-0001-9123-6008}
\affiliation{%
  \institution{University College London (UCL)}
  \city{London}
  \country{United Kingdom}
}
\email{r.moussa@ucl.ac.uk}

\author{Federica Sarro}
\orcid{0000-0002-9146-442X}
\affiliation{%
  \institution{University College London (UCL)}
  \city{London}
  \country{United Kingdom}
}
\email{f.sarro@ucl.ac.uk}

\renewcommand{\shortauthors}{d'Alosio et al.}

\begin{abstract}
    \textbf{Background:} The wide adoption of AI- and ML-based systems in sensitive domains raises severe concerns about their fairness. Many methods have been proposed in the literature to enhance software fairness. However, the majority behave as a black-box, not allowing stakeholders to prioritise fairness or effectiveness (i.e., prediction correctness) based on their needs. 
    \textbf{Aims:} In this paper, we introduce \tool, a novel approach based on multi-objective evolutionary search to optimise fairness and effectiveness in classification tasks. \tool uses a Random Forest (RF) model as a base classifier and searches for the best hyperparameter configurations and data mutation to maximise fairness and effectiveness. Eventually, it returns a set of Pareto optimal solutions, allowing the final stakeholders to choose the best one based on their needs.
    \textbf{Method:} We conduct an extensive empirical evaluation of \tool against 26 different baselines in 11 different scenarios using five effectiveness and three fairness metrics. Additionally, we also include two variations of the fairness metrics for intersectional bias for a total of six definitions analysed.    
   \textbf{Result:} Our results show that \tool can significantly improve the fairness of base classifiers, while maintaining consistent prediction effectiveness. Additionally, \tool provides a more consistent optimisation under all fairness definitions compared to state-of-the-art bias mitigation methods and overcomes the existing state-of-the-art approach for intersectional bias mitigation. 
   \textbf{Conclusions:} \tool is an effective approach for bias mitigation also allowing stakeholders to adapt the development of fair software systems based on their specific needs.
\end{abstract}



\maketitle

\section{Introduction}
\label{sec:Introduction}

With AI- and ML-based software systems widely employed in sensitive domains such as healthcare \cite{canali_challenges_2022}, finance \cite{kozodoi_fairness_2022}, and education \cite{austin_will_2016}, it is critical to ensure that they act in an \textit{unbiased} and \textit{ethical} way. In other words, they must be \emph{fair}. The relevance of \emph{software fairness} has been highlighted not only in research literature, but also in regulations such as the European Union's recently introduced AI Act \cite{noauthor_eu_2023}.

In the software engineering (SE) domain, fairness can be seen as an important, non-functional property of software systems~\cite{zhang2020machine}, and a fairness violation can be defined as a system bug~\cite{chen2024fairness}.
As such, fairness has gained considerable attention from the SE community in recent years, covering the bias assessment and mitigation of AI- and ML-based systems~\cite{chen_diversity_2024,chen_fairness_2024,sarro2023search,hort_bias_2023}. As highlighted in previous studies \cite{daloisio_how_2024,daloisio_democratizing_2023,d2025manila}, the assessment and development workflow of fair software systems is a complex process involving multiple stakeholders. Given a specific use-case domain, there are the domain (and if needed, the legal) expert(s) who define a specific fairness requirement. Next, there are the software engineers and data scientists who implement the system such that the given fairness requirement is satisfied~\cite{daloisio_how_2024,daloisio_democratizing_2023,d2025manila}.

To ensure that a system is fair, data scientists and software engineers can use bias mitigation methods to address and mitigate possible fairness bugs~\cite{chen2023comprehensive,hort_bias_2023}. Many bias mitigation methods have been proposed by the SE and AI communities in recent years~\cite{hort_bias_2023}. However, the majority of them behave like black-box approaches and do not allow stakeholders to choose different solutions based on the different priorities they may have on fairness and prediction correctness (i.e., effectiveness). Indeed, there could be use cases where a domain expert may want to prioritise effectiveness over fairness, like in safety-critical systems.

To overcome this limitation, we can formulate the bias mitigation task as a multi-objective problem that balances fairness and effectiveness \cite{Sarro19Keynote}. The solutions to this problem are multiple, each providing an optimal trade-off between fairness and effectiveness, forming a so-called Pareto front (i.e., a set of solutions in which no solution is strictly better than any other \cite{Harman2011}). Search-based Software Engineering (SBSE) has been proven to be a widely effective and efficient approach to address these kinds of problems \cite{Harman2011,sarro2023search,Sarro19Keynote,daloisio_sustaindiffusion_2025,gong_greenstableyolo_2024,gong_2025}. In SBSE, an SE problem (in our case, fairness bug fixing \cite{chen2023comprehensive}) is formulated as a search task, where optimal solutions are identified within a space of solution candidates through one or more fitness functions that guide the search \cite{Harman2011}.

\begin{figure}[tb]
    \centering
    \includegraphics[width=.6\linewidth]{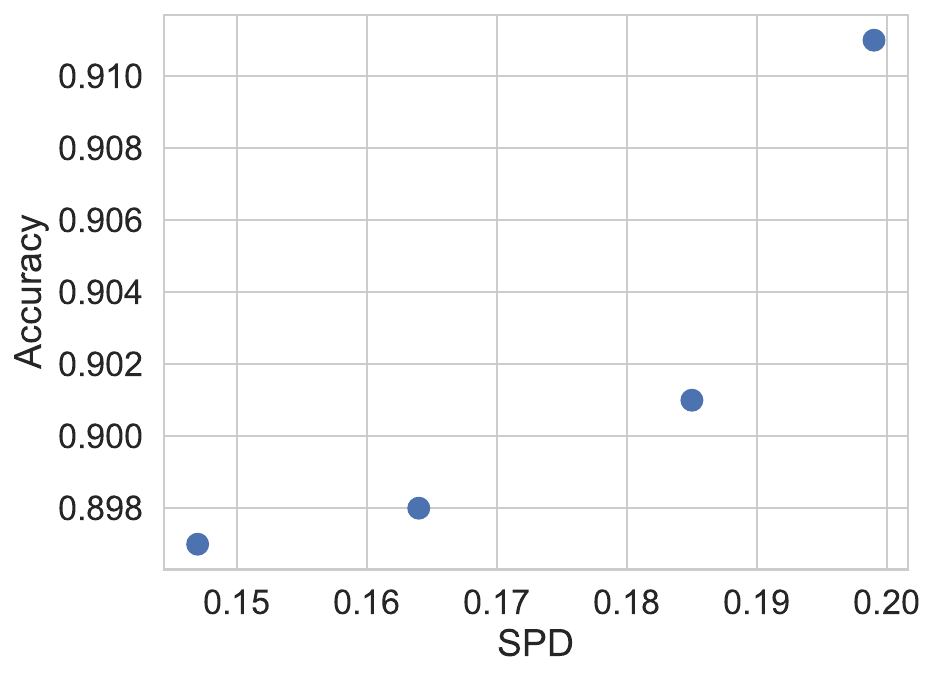}
    \caption{Example of Pareto front solutions returned by \tool. The more solutions are closer to the upper-left corner of the plot, the better.}
    \label{fig:pareto}
\end{figure}

Following the SBSE approach, in this paper, we propose \tool, a Multi-Objective Evolutionary Algorithm (MOEA) that optimises fairness and effectiveness in classification tasks. In particular, \tool uses a Random Forest (RF) model as a base classifier and searches for the best hyperparameter configuration and input data mutation to maximise fairness and effectiveness (more details are reported in Section \ref{sec:method}).

Figure \ref{fig:pareto} reports an example output of \tool for a finance use case \cite{moro2014data}. Each point in the figure represents a solution returned by \tool (i.e., a set of RF hyperparameter values and a data mutation). The $x$-axis in the plot reports the fairness score (in this case, Statistical Parity Difference \cite{hardt2016equality}, with optimal value equal to zero), while the $y$-axis shows the effectiveness score (in this case, Accuracy \cite{rosenfield_coefficient_1986}, with optimal value equal to one) obtained by each solution. The closer solutions are to the upper-left corner of the plot, the better. The set of solutions returned by \tool forms a Pareto front, where no solution is strictly dominated by the others. Therefore, the domain expert can choose the solution that better satisfies their requirements, either maximising fairness, effectiveness, or a compromise between the two. 

We perform an extensive empirical evaluation of \tool against 26 different baselines (including base classifiers, \tool algorithm variations, and state-of-the-art bias mitigation methods), using 11 different scenarios (i.e., datasets and sensitive attributes combinations) under five effectiveness metrics and three fairness definitions. We demonstrate how \tool significantly improves the fairness of base classifiers, while maintaining consistent prediction effectiveness. Moreover, \tool provides a more consistent optimisation under all fairness definitions compared to state-of-the-art bias mitigation methods. Finally, we show how \tool overcomes the existing state-of-the-art approach for intersectional bias mitigation. 

To summarise, the main contributions of our work are: 

\noindent\textit{i)} A novel multi-objective evolutionary algorithm (\tool) for fairness and effectiveness improvement employing both one and multiple sensitive variables;

\noindent\textit{ii)} An extensive empirical evaluation of \tool against 26 baselines in 11 different scenarios;
    
\noindent\textit{iii)} A discussion on the practical implications of adopting \tool;

\noindent\textit{iv)} A replication package providing the Python implementation of \tool and the results of our evaluation \cite{replpackage,d_aloisio_2025_17879088};

\section{Background and Related Work}
\label{sec:Background}

In the following, we present the main concepts and discuss related work in the context of software fairness.


\subsection{Software Fairness}

Software \emph{fairness} is defined as: \textit{"The absence of prejudice and favouritism of a software system toward individuals or groups"}~\cite{mehrabi_survey_2021}. When a system behaves unfairly, it is said to be \emph{biased}. Software fairness is mainly studied in the context of AI-based systems, with the aim of assessing and mitigating the bias learned by AI and ML components~\cite{mehrabi_survey_2021,daloisio_how_2024,hort_bias_2023}. In this context, bias can originate from three main sources~\cite{mehrabi_survey_2021}: the \textbf{data} used to train the AI and ML components, a \textbf{biased implementation} of the AI and ML components, and the \textbf{people} that interact with those components. In this paper, we focus on the first source of bias, aiming to mitigate the bias of an ML model when it is trained on biased data. More specifically, we focus on the task of ML classification with structured, tabular data, which is the main task investigated in the software fairness domain~\cite{hort_bias_2023,mehrabi_survey_2021}. 

In general, the fairness of a system can be assessed following two main criteria: \textbf{individual} and \textbf{group} fairness~\cite{speicher2018unified}. 
\textbf{Individual} fairness requires that two individuals who are similar to one another receive the same treatment (i.e., an ML model should make identical predictions). 
Most of the time, two individuals are treated as similar if they only differ in sensitive attributes\footnote{In the rest of this paper, we will use the terms "sensitive attributes", "protected attributes" or "sensitive features" as synonyms.} (e.g., \textit{ethnicity}, \textit{gender}, \textit{age}).
\textbf{Group} fairness, on the other hand, addresses fairness by treating population groups, defined by protected attributes (like \textit{ethnicity}, \textit{gender}, or \textit{age}), equally. In this work, we focus on the group fairness criteria, as they are more common and have been more extensively addressed by previous works \cite{hort_bias_2023,daloisio_how_2024}. Specifically, many group fairness definitions and corresponding metrics have been proposed in the literature \cite{mehrabi_survey_2021,caton_fairness_2023}. The general idea behind all group fairness definitions is that, given two groups named \textbf{privileged} and \textbf{unprivileged} (e.g., \textit{men} and \textit{women}), they must have the same probability of having a given \textbf{positive} outcome from the ML model, possibly conditioned on the ground truth label. For instance, following the \textbf{Statistical Parity} group fairness definition \cite{hardt2016equality}, \textit{men} and \textit{women} should have the same probability of being admitted to a university \cite{austin_will_2016}. In this paper, we consider the three most adopted group fairness definitions \cite{hort_bias_2023}, namely \textbf{Statistical Parity}, \textbf{Equal Opportunity}, and \textbf{Average Odds} (see Sections \ref{sec:ftiness} and \ref{sec:metrics}).

In addition to determining the degree of bias of an ML model, one can set out to reduce it through the use of bias mitigation methods. Generally, improvement in fairness implies a reduction in the effectiveness of a model's predictions \cite{chen2023comprehensive,hort_bias_2023,hort2021fairea}, and all bias mitigation methods try to identify the optimal trade-off between fairness and effectiveness.
In particular, there are three types of bias mitigation methods based on when they are applied in an ML workflow: \textbf{pre-processing}, \textbf{in-processing}, and \textbf{post-processing}~\cite{friedler2019comparative,hort_bias_2023,mehrabi_survey_2021}. 
\textbf{Pre-processing} bias mitigation methods aim to reduce bias by applying changes to the training data.
For instance, one can assign more weight to data instances for a population group that is prone to being misclassified~\cite{kamiran2012data,calders2009building,daloisio_debiaser_2023}.
\textbf{In-processing} bias mitigation methods make changes to the design and training process of ML models to achieve fairness. One example is the inclusion of fairness metrics as part of the training loss~\cite{beutel2019putting}. 
Alternatives include the tuning of hyperparameters~\cite{tizpaz2022fairness} or the use of ensembles, where each model can consider different population groups~\cite{dwork2018decoupled} or metrics~\cite{chen2022maat}. 
\textbf{Post-processing} bias mitigation methods are applied once an ML model has been successfully trained. This can involve changes to the prediction made by a model~\cite{hardt2016equality} or modifying the model itself~\cite{savani2020intra,kamiran2010discrimination}. 
We discuss the bias mitigation methods employed as baselines in our study in Section \ref{sec:ML Algorithms}.

\subsection{Related Work}\label{sec:Related Work}

Software fairness gained considerable relevance in the SE community in recent years. In this context, bias assessment and mitigation could be considered as fixing a fairness bug in the system and satisfying a given fairness requirement~\cite{chen2024fairness}. 

As outlined in previous work \cite{hort_bias_2023,daloisio_how_2024}, recent years have seen the emergence of various methods in the fields of Software Engineering (SE) and Machine Learning (ML) aimed at mitigating bias across different processing levels. These methods focus on sensitive groups identified by a single sensitive variable~\cite{kamiran2012data,hardt2016equality,chen2022maat,daloisio_debiaser_2023,chen_diversity_2024,dwork2018decoupled} as well as  sensitive groups classified by more than one sensitive variable (a.k.a. intersectional fairness) ~\cite{chen_diversity_2024,chen2022maat,chakraborty2021bias,daloisio_debiaser_2023}. 

Most of the proposed approaches operate as black-box models, primarily focusing on bias mitigation. However, earlier studies in SE have emphasised that the process of assessing and mitigating bias is complex and involves multiple stakeholders, including decision makers, domain experts, software engineers, and data scientists. Decision makers and domain expert define bias within a specific domain by identifying privileged and unprivileged groups and establishing what constitutes a positive outcome. They also determine how bias can be measured and addressed~\cite{daloisio_how_2024,daloisio_democratizing_2023}. 
Therefore, in some instances, stakeholders may prioritise improving overall effectiveness over bias mitigation or may want to minimise specific classification errors, such as false positives. With \tool, we aim to address this need by leveraging the flexibility and adaptability of Multi-Objective Evolutionary Algorithms (MOEAs), which fit well for these kinds of tasks \cite{sarro2023search}.

Prior studies have already explored MOEAs for bias mitigation and to improve the effectiveness of ML models in classification tasks. Concerning bias mitigation, Hort et al. applied MOEAs as a post-processing approach to mutate the predictions of Logistic Regression and Decision Tree classifiers \cite{hort_search-based_2024}. Similarly, Hort et al. also applied MOEAs as an in-processing approach for bias mitigation in word embeddings \cite{hort_multi-objective_2023}. Perera et al. employed a search-based approach for fairness testing of regression systems in the health domain \cite{perera_search-based_2022}. Unlike previous studies, \tool employs a combined pre- and in-processing approach, utilising an MOEA to mutate the input dataset and tune the hyperparameters of the RF model to maximise both fairness and effectiveness. Concerning effectiveness improvement, Moussa et al. employed MOEAs to generate ensemble models for defect prediction \cite{moussa_meg_2022}. Similarly, other studies from the ML domain employed MOEAs for several effectiveness optimisation techniques, such as feature selection \cite{Chen_2018,Han_2021} or hyperparameter optimisation \cite{Dos_Santos_2006,Fletcher_2020}.

Finally, the choice to mutate the sensitive features of the input dataset to improve fairness is driven by the previous work of Chen et al. on intersectional bias mitigation \cite{chen_diversity_2024}. In their work, the authors generate an ensemble of different classifiers, each one trained on a different version of the input dataset, such that all possible combinations of the sensitive features are covered. In our work, we employ a different approach by including the mutation of the dataset as part of the search. Nevertheless, we include the work of Chen et al. (FairHOME) as one of the baselines for our evaluation (see Section \ref{sec:ML Algorithms}).

\section{Methodology}
\label{sec:method}

\begin{figure}[tb]
  \includegraphics[width=1\linewidth]{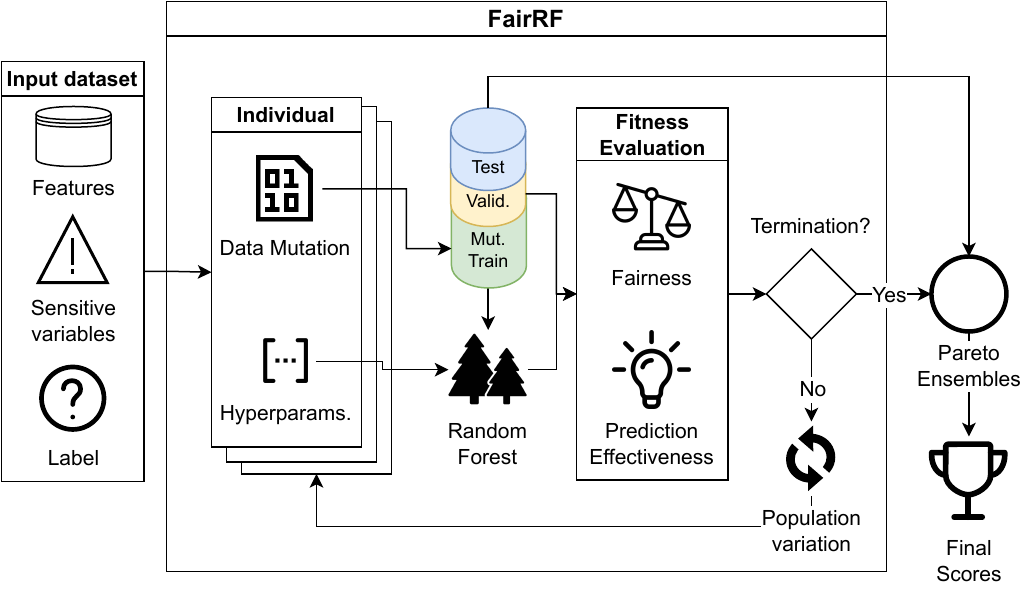}
    \caption{Overview of the \tool approach.}
  \label{fig:meg}
\end{figure}

The goal of our study is to identify a classification approach that optimises fairness (both single and intersectional) and effectiveness, while simultaneously allowing users to prioritise either fairness or effectiveness, depending on the use case. To achieve this, we propose \tool, a multi-objective evolutionary algorithm (MOEA) that automatically searches for optimal configurations of a Random Forest (RF) classifier and data mutations, taking both fairness and effectiveness in account. The choice for an RF model as a base classifier is driven by its effective application in previous fairness studies \cite{hort_bias_2023,chen_diversity_2024} and by empirically evaluating different base classifiers on different datasets (see the answer to RQ1 in Section \ref{sec:Results:RQ1})

Figure~\ref{fig:meg} depicts an overview of \tool. The input of \tool is a tabular dataset that includes a set of non-sensitive features, one or more sensitive variables, and a label column. Then, the evolutionary process begins. First, an initial population of unique individuals is randomly generated. Each individual is modelled as a dictionary of: a \textit{Data Mutation Value} that indicates the percentage of sensitive attributes in the training dataset that must be mutated, and a set of hyperparameters for the RF model. More details about the individuals' representation are reported in Section~\ref{sec:representation}. The dataset is then split into training (50\%), testing (30\%) and validation (20\%). The training set is first mutated following the \textit{Data Mutation Value} and then used to train the RF model, with the given hyperparameter configurations. The validation set is instead used to compute the fitness functions for a given individual (more details in Section \ref{sec:ftiness}). If the termination criteria are not met (i.e., not all generations are explored), then the best individuals are selected and the population evolves using the genetic operators described in Section \ref{sec:operators}. When all generations are explored, the solutions providing the best fairness-accuracy trade-offs are trained on the training plus validation sets (80\%) and evaluated on the unseen testing set to compute the final scores. 

In the following, we describe \tool in detail by discussing its main components: the individuals, the fitness functions, and the genetic operators \cite{sarro_multi-objective_2016,moussa_meg_2022}.

\subsection{Representation}
\label{sec:representation}

\begin{table}[tb]
    \centering
    \caption{Individual Representation.}
    \label{tab:individual}
    \resizebox{.7\linewidth}{!}{\begin{tabular}{l|r}
    \toprule
      \textbf{Attribute} & \textbf{Value Range} \\
      \midrule
      \midrule
       Data Mutation Value & \{0.1 : 1.0\} with a step of 0.1 \\
    \midrule
       Estimators Num. & \{10, 20, 50, 80, 100, 150, 200\} \\
       Quality Criterion & \{\textit{gini}, \textit{entropy}, \textit{log loss}\} \\
       Max Depth & \{\textit{None}, 10, 15, 20, 30, 40, 50\} \\
       Min Samples Split & \{2, 3, 4\}\\
       Max Features & \{\textit{sqrt}, \textit{log2}, \textit{None}\}\\
       \bottomrule
    \end{tabular}}
\end{table}

Each individual in \tool is represented as a dictionary of a \textit{data mutation value} and RF hyperparameters. Table \ref{tab:individual} summarises the individual representation.

The \textit{Data Mutation Value} is modelled as a float that ranges from 0.1 to 1.0 and represents the percentage of sensitive features to mutate in the input dataset. Specifically, previous research has shown how diversity in the sensitive variables of the input dataset increases fairness in the model~\cite{chen_diversity_2024}. Therefore, given a value $n$ of the \textit{Data Mutation Value} for a given individual, \tool mutates a randomly selected $n\%$ of the sensitive variables in the training set. Following~\cite{chen_diversity_2024}, \tool performs a bit-flip mutation of the sensitive features, applying the function $f(x) =1-x$, where $x$ is the original value of the sensitive feature.\footnote{Following several fairness studies \cite{chen_diversity_2024,daloisio_debiaser_2023,hort_search-based_2024,chen_fairness_2024}, we assume that sensitive features are encoded as binary values} We would like to remark how this mutation is only applied on the training set, while the testing set is kept unchanged to not bias the final results.

In addition to the \textit{Data Mutation Value}, each individual includes the primary RF hyperparameters. Specifically, following previous studies \cite{moussa_meg_2022}, we consider the following: \ding{228}~\textit{Number of estimators}: this parameter specifies the number of trees in the forest. The default value is 100, and we also explore the values  \{10, 20, 50, 80, 150, 200\}; \ding{228}~\textit{Quality Criterion}: this parameter highlights the function to measure the quality of splits in the tree. The default function is the Gini Impurity. In \tool, we also explore Entropy and Logistic Loss \cite{breiman2017classification}; \ding{228}~\textit{Maximum Depth}: this parameter specifies the maximum depth of a tree. Default value is \textit{None} (i.e., no maximum depth). In this work, we also explore the values \{10, 15, 20, 30, 40, 50\}; \ding{228}~\textit{Minimum Samples Split}: this parameter specifies the minimum number of samples required to split an internal node in the tree. The default value is 2. In \tool, we also explore the values 3 and 4. \ding{228}~\textit{Maximum Features}: This parameter specifies the function to count the maximum number of features when identifying the best split. The default function is \textit{square root}. In this work, we also explore \textit{log2} and \textit{no function}.

\subsection{Fitness Functions}\label{sec:ftiness}

After training an RF model on the mutated training set using the hyperparameters of a specific individual, the fitness of that individual is evaluated by calculating both fairness and effectiveness scores of the RF model on the testing set. Both fitness functions are given the same weight of 1. In the following, we describe them in detail.

As highlighted in Section \ref{sec:Background}, there are many definitions of fairness available in the literature \cite{mehrabi_survey_2021,caton_fairness_2023,hort_bias_2023}. In \tool, we use \textbf{Statistical Parity Difference (SPD)} as the fitness function to evaluate the fairness of an individual. SPD calculates the difference in probabilities for individuals from the unprivileged ($A=0$) and privileged ($A=1$) groups to receive a positive prediction~\cite{dwork2012fairness}:

\begin{equation}\label{eq:spd}
\text{SPD} =  Pr(\hat{y} = 1 | A=0)\\
- Pr(\hat{y} = 1 | A=1)
\end{equation}

Following previous studies \cite{chen_diversity_2024,hort_bias_2023,daloisio_debiaser_2023}, we consider absolute values of this metric. Therefore, a value of 0 for this metric highlights fairness, while a value of 1 highlights complete bias. Thus, this fitness function is \textit{minimised} by \tool.

Similar to fairness, many definitions and metrics have been proposed to measure the effectiveness of a model's predictions. In \tool, we employ \textbf{Accuracy} as the fitness function to measure the prediction's effectiveness. \textbf{Accuracy} is a widely adopted metric in classification tasks, which computes the proportion of true positive (TP) and true negative (TN) predictions over the whole model's predictions \cite{rosenfield_coefficient_1986}:\footnote{In the following, FP means false positives, while FN means false negatives.}
\begin{equation}
    \text{Accuracy} = \frac{TP+TN}{TP+TN+FP+FN}
\end{equation}
This metric ranges from 0 to 1, where 1 highlights perfect correctness in the model's predictions. Therefore, this metric is \textit{maximised} by \tool.

We did not include additional fairness and effectiveness definitions as fitness functions to avoid complicating the overall search process \cite{Harman2011}. Future research can investigate the impact of employing different fairness and effectiveness definitions as fitness functions. However, as we show in Section \ref{sec:Results}, the fitness functions employed in \tool yield solutions that provide significantly better results even under definitions that are not directly optimised as fitness functions.

\subsection{Genetic Operators}\label{sec:operators}


\tool employs NSGA2 as MOEA implementation \cite{deb2002nsga2}. NSGA2 is a very popular global search algorithm and has been shown to be widely effective and efficient for many multi-objective optimisation problems \cite{Harman2011}. It follows the general behaviour of genetic evolutionary algorithms, where an initial population progressively evolves following Darwin's theory of biological evolution \cite{Harman2011}. First, a population of randomly generated individuals is created. The quality of each individual is assessed using the fitness functions defined in Section \ref{sec:ftiness}. Individuals are then sorted using a non-dominated sorting strategy, and the best individuals go to the evolution process. Each selected individual goes through crossover and mutation operations with a given probability. The fitness of the newly generated individuals is again evaluated, and the best ones are passed onto the next generation. The evolution process continues until a maximum number of generations is reached. 

For crossover, \tool uses a Single Point Crossover operator with 60\% probability.
This crossover operator takes two parents and combines their values to generate two children. First, it chooses an index from the parents' dictionary at random based on which it splits the parents into two parts, the left part and the right one. It then combines the left part of parent one with the right part of parent two to generate the first child, and the right part of parent one with the left part of parent two to create the second child.

After the crossover, the children undergo mutation with a probability of 20\%. \tool uses a Simple Random Mutation operator. This operator replaces each attribute of the individual with those of a newly generated individual, using an inner mutation probability of 15\% (1/6). It means that, if an individual undergoes the mutation, it is expected that at least one of its attributes will be mutated.  


In our experiments, we set the population size to 50 and the stopping condition to 25 generations. The selection rate (i.e., the number of individuals to select for the next generation) is set to 5. Ultimately, all non-dominated solutions are returned.

\subsection{Implementation Aspects}

\tool has been implemented in Python 3.9.23. The MOEA has been implemented using the widely adopted open-source \texttt{deap} library, which provides implementations of NSGA2 as well as crossover and mutation operations \cite{DEAP_JMLR2012}. The implementation of the RF model, as well as the Accuracy fitness function, is taken from the \texttt{scikit-learn} Python library \cite{pedregosa_scikit-learn_2011}. The implementation of the SPD fitness metric is taken from the widely adopted \texttt{IBM AIF360} library \cite{bellamy_ai_2019}. 

All experiments were conducted on a CentOS machine equipped with 32 Intel(R) Xeon(R) Gold 6140M CPUs. A complete run of the algorithm (i.e., 25 generations) takes on average $\sim$10 minutes on this machine. However, this result may vary based on the input dataset.

\section{Empirical Evaluation}\label{sec:Experimental}
To validate the effectiveness of \tool, we benchmark it against variations of the proposed approach that employ different base classifiers, ML base classifiers, and the state-of-the-art (SOTA) bias mitigation methods for single and intersectional fairness. Specifically, we aim at answering four research questions (RQs) as detailed below.

\subsection{Research Questions}

\begin{rqs}

\item \textbf{\rqtwoshort} \textit{\rqtwo} The first RQ aims to assess whether the choice of adopting an RF model as a base classifier in our approach is the best solution compared to other base classification methods. Once we demonstrate that \tool overcomes other variations of the algorithm that employ different base classifiers, we can benchmark it against other baselines.
    
\item  \textbf{\rqoneshort:} \textit{\rqone} ML base classifiers have been widely proposed and considered as benchmarks in previous fairness studies  \cite{daloisio_debiaser_2023,hort_bias_2023,chen_diversity_2024}. Therefore, this research question investigates and compares the fairness and effectiveness of \tool with that of traditional ML base classifiers and a simple Random Search (RS) approach. We consider this a ``sanity check'' given that any newly proposed model that cannot generally outperform base classifiers cannot be considered a scientific advancement in the SOTA. 
    
\item \textbf{\rqthreeshort} \textit{\rqthree} Several methods have been proposed in recent literature to mitigate bias with a single sensitive variable at different processing levels \cite{hort_bias_2023}. Therefore, in this RQ, we benchmark \tool against the main methods that have been used in the literature \cite{hort_bias_2023}.    
    
\item \textbf{\rqfourshort} \textit{\rqfour} Finally, recent research highlighted the need to address intersectional fairness \cite{chen_fairness_2024}. In this perspective, we benchmark \tool against the FairHOME approach, which has been shown to outperform other approaches proposed for intersectional fairness \cite{chen_diversity_2024}.
\end{rqs}

\subsection{Benchmark Techniques}
\label{sec:ML Algorithms}

\subsubsection{Base Classifiers}

To answer \textbf{RQ1}, we benchmark \tool against four variations of the algorithm. Each variation of the algorithm performs the same mutation of the input dataset as \tool, but employs and fine-tunes different base classifiers. The choice of these base classifiers is driven by their widespread adoption in classification tasks and previous fairness studies \cite{hort_bias_2023}, as well as their relatively small computational complexity.
The algorithms analysed are reported in the following (for each hyperparameter, we report in brackets the explored values):

\noindent\ding{228} \textbf{FairLR}: This algorithm uses a Logistic Regression model as the base classifier and tunes the following hyperparameters: \textit{Fit Intercept} (\textit{True}, \textit{False}), \textit{Class Weight} (\textit{No weight}, \textit{Balanced}), \textit{Solver} (\textit{lbfgs}, \textit{liblinear}, \textit{newton-cg}, \textit{newton-cholesky}, \textit{sag}, \textit{saga}).

\noindent\ding{228} \textbf{FairKNN}: This algorithm uses a K-Nearest Neighbours classifier and tunes the following hyperparameters: \textit{Number of neighbours} (\{2, 3, 4, 5, 6, 8, 10, 12, 14, 18, 20\}), \textit{Weight} (\textit{uniform}, \textit{distance}), \textit{Algorithm} (\textit{auto}, \textit{ball tree}, \textit{kd tree}, \textit{brute}), \textit{P} (\{1, 2\}).

\noindent\ding{228} \textbf{FairCART}: This algorithm uses a single Classification and Regression Tree (CART) as a base classifier and tunes the following hyperparameters: \textit{Criterion} (\textit{gini}, \textit{entropy}, \textit{log loss}), \textit{Maximum depth} (\{[\textit{No depth}, 10, 15, 20, 30, 40, 50\}), \textit{Splitter} (\textit{best}, \textit{random}), \textit{Maximum Features} (\textit{sqrt}, \textit{log2}, \textit{No function}).

\noindent\ding{228} \textbf{FairSVM}: This algorithm employs a Support Vector Machine classifier and tunes the following hyperparameters: \textit{C} (\{0.1, 1.0, 10.0, 100.0\}), \textit{Kernel} (\textit{linear}, \textit{poly}, \textit{rbf}, \textit{sigmoid}), \textit{Degree} (\{2, 3, 4\}), \textit{Gamma} (\textit{scale}, \textit{auto}). 

To answer \textbf{RQ2}, we benchmarked \tool against the same base classifiers employed for \textbf{RQ1}, using their implementation from the \texttt{scikit-learn} library \cite{pedregosa_scikit-learn_2011}. For each classifier, we tune its hyperparameters by exploring the same search space as \textbf{RQ1} using a Grid Search approach \cite{Harman2011}, but we keep the training dataset unchanged. In addition, we also benchmark \tool against a Random Search (RS) algorithm, which, at each iteration, randomly chooses and returns an individual from the search space described in Section \ref{sec:representation}. Following research standards \cite{Harman2011}, the number of individuals evaluated by the RS algorithm in each iteration is equal to the average number of evaluations performed by \tool in each generation (i.e., six).




\subsubsection{State-Of-The-Art Bias Mitigation Methods} 

We benchmark \tool against the following methods to answer \textbf{RQ3} and \textbf{RQ4}. These methods are the SOTA in terms of bias mitigation for single and intersectional fairness \cite{hort_bias_2023}. 

In particular, concerning \textbf{RQ3}, we consider the following methods: 

\noindent\ding{228}~\textbf{Reweighing (RW)}~\cite{kamiran2012data} is a pre-processing method which adjusts the weight groups in the data based on group and label. 

\noindent\ding{228}~\textbf{Adversarial Debiasing (ADV)}~\cite{zhang2018mitigating} is an in-processing method which trains a classifier for achieving high accuracy while simultaneously reducing the ability to predict protected attributes from internal representations. 
Thereby, predictions can be made without knowledge of sensitive information. 

\noindent\ding{228}~\textbf{MAAT}~\cite{chen2022maat} is an in-processing method which proposes the use of ensembles which combine ML models trained for different objectives: fairness and performance. The goal is to achieve good trade-offs between fairness and performance. 
Specifically, MAAT trains two models, one aimed at fairness and one aimed at performance, and averages their predictions. 

\noindent\ding{228}~\textbf{Debiaser for Multiple Variables (DEMV)}~\cite{daloisio_debiaser_2023} is a variation of the pre-processing FairSMOTE algorithm~\cite{chakraborty2021bias} that improves fairness by adding or removing items from each sensitive group through random sampling. Different from FairSMOTE, DEMV adds new items by randomly replicating existing items in the dataset.

\noindent\ding{228}~\textbf{Equalized Odds Post-processing (EOP)}~\cite{hardt2016equality,pleiss2017fairness}
modifies the predicted labels such that the equalized odds difference becomes minimised. 

Concerning \textbf{RQ4}, we benchmark \tool against the \textbf{FairHOME} approach \cite{chen_diversity_2024}. This algorithm generates different mutations of the input dataset such that all intersectional groups are equally represented. Then, it trains an ensemble of classifiers, where each single classifier is trained on a mutated version of the dataset. This method has been recently presented and overcomes all other methods proposed for intersectional fairness \cite{chen_diversity_2024}. 

 We employ the implementation of RW, AOD, and EOP from the \texttt{AIF360} library~\cite{bellamy_ai_2019}, while we employ the implementations of MAAT, DEMV, and FairHOME provided in their respective replication packages.

\subsection{Datasets}
\label{sec:datasets}

We employ the following datasets in our evaluation, which are widely adopted in the fairness literature \cite{d2025towards} and are summarised in Table \ref{tab:data}:

\begin{table}[tb]
    \centering
    \caption{Dataset details.}
    \label{tab:data}
\resizebox{.7\linewidth}{!}{\begin{tabular}{lrrr}
\toprule
\textbf{Name} & \textbf{Sens. Attribute(s)} & \textbf{Features} & \textbf{Instances} \\ \midrule\midrule
Adult & \textit{race}, \textit{sex} & 99 & 45,222 \\
Compas & \textit{race}, \textit{sex} & 402 & 6,172 \\
German & \textit{sex}, \textit{age} & 59 & 1,000 \\
Bank & \textit{age} & 58 & 30,488 \\
MEPS & \textit{race} & 139 & 15,830 \\ \bottomrule
\end{tabular}}
\end{table}

\noindent\ding{228} The Adult Income dataset (\textbf{Adult})~\cite{adultdata} provides a collection of information about individuals from the 1994 U.S. census. The binary prediction task is to determine whether an individual has a yearly income above 50 thousand USD.

\noindent\ding{228} The ProPublica Recidivism dataset (\textbf{Compas},  \textit{Correctional Offender Management Profiling for Alternative Sanctions}~\cite{compas}) contains details about the criminal history of offenders in Broward County, Florida.
The classification task is to determine whether an individual is to recidivate within the next two years.

\noindent\ding{228} The Bank Marketing dataset (\textbf{Bank})~\cite{moro2014data}  describes a marketing campaign performed by a Portuguese banking institution, where the classification task is to determine whether potential customers will subscribe to term deposits. Here, individuals are divided into young and old at a threshold age of 25. 

\noindent\ding{228} The German Credit dataset (\textbf{German})~\cite{germandata} is the smallest included dataset, with 1000 instances. Individuals are labelled as having a good or bad credit risk.

\noindent\ding{228} The Medical Expenditure Panel Survey (\textbf{MEPS}) contains information about the utilisation of medical providers in 2019. The goal is to predict whether furniture is utilised or not.\footnote{https://meps.ahrq.gov/mepsweb/}

For the first three \textbf{RQs} we employ all the datasets, while for \textbf{RQ4}, we use \textbf{Adult}, \textbf{Compas}, and \textbf{German} since they provide two sensitive attributes.

\subsection{Evaluation Metrics and Methods}
\label{sec:measures}

\subsubsection{Metrics}\label{sec:metrics}

To comprehensively evaluate \tool, we compute additional effectiveness and fairness metrics for each solution returned, in addition to those directly optimised as fitness functions (Accuracy and SPD, see Section~\ref{sec:ftiness}).

Concerning effectiveness, we include the following metrics: \ding{228}~\textbf{Precision:} defined as the ability of the model in identifying true positives~\cite{buckland_relationship_1994};
\ding{228}~\textbf{Recall:} defined as the ability of the model in identifying all the true positives~\cite{buckland_relationship_1994};
\ding{228}~\textbf{F1 Score:} defined as the harmonic mean between Precision and Recall~\cite{buckland_relationship_1994};
\ding{228}~\textbf{MCC:} the \textit{Matthews Correlation Coefficient}, which measures the correlation between the true and predicted label~\cite{moussa2022use,Menzies2007Response}. Precision, Recall, and F1 Score range from 0 to 1, where 1 is the optimal value. On the contrary, MCC ranges from -1 to 1, with 1 still being the optimal value.

Concerning fairness, we complement the SPD metric with two other widely adopted fairness definitions \cite{mehrabi_survey_2021}: 

\noindent\ding{228}~\textbf{Equal Opportunity Difference (EOD)}: is calculated as the difference in True Positive Rate (TPR) of the privileged and unprivileged group~\cite{hardt2016equality}:
\begin{equation}
      \text{EOD}=  TPR_{A=unprivileged} - TPR_{A=privileged}    
\end{equation}

\noindent\ding{228}~\textbf{Average Odds Difference (AOD)} averages the difference in TPR as well as False Positive Rate (FPR)~\cite{hardt2016equality}:
\begin{equation}
\begin{split}
      \text{AOD}=  \frac{1}{2}((FPR_{A=unprivileged} - FPR_{A=privileged})\\
           + (TPR_{A=unprivileged} - TPR_{A=privileged}))
    \end{split}
\end{equation}
Those metrics consider sensitive groups identified by a single sensitive attribute. Therefore, following previous studies \cite{chen_diversity_2024,chen_fairness_2024}, in the answer to \textbf{RQ4}, we consider the following extensions to measure intersectional fairness: 

\noindent\ding{228}~\textbf{Worst-Case-Scenario (WCS)}: computes SPD, EOD, and AOD as the difference between subgroups ($s \in S$) with maximum and minimum discrimination. For instance, WCS-SPD is defined as:
\begin{equation}
    \text{WCS-SPD} = max_{s \in S}(Pr[\hat{y} = 1 | S = s]) - min_{s \in S}(Pr[\hat{y} = 1 | S = s])
\end{equation}

\noindent\ding{228}~\textbf{Average (AVG)}: computes SPD, EOD, and AOD as the average between the scores obtained for each subgroup ($s \in S$). For instance, AVG-SPD is defined as:
\begin{equation}
    \text{AVG-SPD} = \frac{\sum_{s \in S} Pr(\hat{y} = 1 | S = s)}{|S|}
\end{equation}

\subsubsection{Methods}

To address the stochastic behaviour of MOEA and ML models, we repeat each experiment 20 times, using a different training and testing split of the input dataset each time. The same approach has been followed for all baseline methods, and we ensured that the same training and testing split as \tool was used by fixing the random seed. Additionally, when comparing \tool against baselines that return a single solution, we average the scores of the Pareto solutions returned by \tool, such that we have a set of 20 scores for each evaluated approach.

To address the statistical significance of the obtained results, we employed the non-parametric \textit{Wilcoxon Sign-Ranked} test \cite{wilcoxon1992individual}. The Wilcoxon test verifies the null hypothesis that the mean of two dependent samples is equal and raises the bar for significance by making no assumption about the underlying distribution \cite{Arcuri2014}. In particular, the null hypothesis that we check is \textit{"$H_0:$ The objective $O$ obtained by \tool is not improved with respect to the baseline approach $x$"}. The alternative hypothesis is: \textit{"$H_1:$ The objective $O$ obtained by \tool is improved with respect to the baseline approach $x$"}. For effectiveness metrics ``\textit{improved}" means that the score obtained by \tool is higher than the baseline. On the contrary, for fairness metrics ``\textit{improved}" means that the score obtained by \tool is lower than the baseline. Following standards \cite{sarro_multi-objective_2016,Arcuri2014}, we set the confidence value to 0.05. Therefore, we reject the null hypothesis if the test's $p\text{-value is} < 0.05$. Moreover, we used effect size to assess whether the statistical significance has a practical significance effect size~\cite{Arcuri2014}. To this end, we use Vargha and Delaney's $\hat{A}_{12}$ non-parametric effect size measure, as it is recommended to use a standardised measure rather than a pooled one like Cohen's $d$ when not all samples are normally distributed~\cite{Arcuri2014}, as in our case. 
The $\hat{A}_{12}$ statistic measures the probability that an algorithm $A$ yields greater values for a given performance measure $M$ than another algorithm $B$. Following previous works \cite{sarro_multi-objective_2016,hort_search-based_2024}, we consider values between $(0.44, 0.56)$ \textit{negligible} differences, values between $[0.56, 0.64)$ and $(0.36, 0.44]$ \textit{small} differences, values between $[0.64, 0.71)$ and $(0.29, 0:44]$ \textit{medium} differences, and values between $[0.0, 0.29]$ and $[0.71, 1.0]$ \textit{large} differences. 

When comparing \tool against SOTA bias mitigation methods, we analyse the trade-off between the metrics in terms of Pareto-optimality \cite{Harman2011}. Pareto-optimality is computed by identifying the Pareto Front from all the solutions returned by all algorithms and counting the number of times a solution from a given algorithm appears in the front \cite{Harman2011}. We recall how the Pareto Front is defined as the set of solutions $ x \in X$ such that $x$ is better than all other solutions in $X$ in at least one objective, and no other solution dominates $x$ in all objectives.

Finally, to address \textbf{RQ1}, since we compare different MOEAs, we also report the Hypervolume score. Hypervolume is a metric used in multi-objective optimisation problems to evaluate the quality of Pareto-optimal solutions in the objective space with respect to a given reference point \cite{10.1145/3453474}. Higher values indicate better coverage of the objective space. Hypervolume has been computed as follows. First, we collected the Pareto Front solutions returned in the 20 runs by each benchmark analysed in \textbf{RQ1}. Following \cite{8637212}, the reference point has been identified by selecting the worst value for each metric analysed among the whole Pareto Fronts, plus an $\epsilon$ value of 0.5. Finally, we compute the Hypervolume of the Pareto Front returned by each approach in the 20 runs. This whole process has been repeated for each dataset + sensitive attribute combination.

\section{Results}
\label{sec:Results}

In this section, we present and discuss the results of all research questions addressed in our work. 

\subsection{Answer to RQ1 --  \tool vs Algorithms Variations}
\label{sec:Results:RQ1}

Table~\ref{tab:hypervolume} reports the mean and standard deviation of Hypervolume scores reported by each model in the 20 runs for each scenario (i.e., dataset -- sensitive attribute combination) analysed. In the table, the highest Hypervolume for each scenario is highlighted in \textbf{bold}. Moreover, we highlight cases where \tool is significantly better in \colorbox{lightblue}{blue} (i.e., Wilcoxon test $p-$value $< 0.05$ and \textit{large} effect size), while we highlight cases where \tool is significantly worse than the use of the respective other models in \colorbox{orange}{orange} (i.e.,  Wilcoxon test $p-$value $> 0.95$).

\begin{table}[tb]
    \centering
    \caption{RQ1: Comparison of Hypervolume scores for each scenario (dataset and sensitive attribute combination). Highest values are highlighted in bold. Winning cases ($p$-value $<$ 0.05 and \textit{large} effect size) are highlighted in \colorbox[HTML]{AAD5D3}{blue}, while losing cases ($p$-value $>$ 0.95) are highlighted in \colorbox{orange}{orange}.}
    \label{tab:hypervolume}
    \resizebox{\linewidth}{!}{
 \begin{tabular}{ll||rrrr|r}
        \toprule
        \textbf{Data} & \textbf{Attr.} & \textbf{FairCART} & \textbf{FairKNN} & \textbf{FairLR} & \textbf{FairSVM} & \textbf{FairRF} \\
        \midrule\midrule
        \multirow[c]{2}{*}{Adult} & race & \isBetter 0.119 $\pm$ 0.004 & \isBetter 0.091 $\pm$ 0.006 & \isBetter 0.108 $\pm$ 0.005 & \isBetter 0.112 $\pm$ 0.006 &\textbf{ 0.124 $\pm$ 0.005} \\
         & sex & 0.232 $\pm$ 0.018 & \isBetter 0.178 $\pm$ 0.01 & \isWorse \textbf{ 0.245 $\pm$ 0.014} & 0.244 $\pm$ 0.011 & 0.23 $\pm$ 0.011 \\
        \cline{1-7}
        \multirow[c]{1}{*}{Bank} & age & 0.071 $\pm$ 0.01 & \isBetter 0.036 $\pm$ 0.004 & \isBetter 0.063 $\pm$ 0.01 & \isBetter 0.07 $\pm$ 0.004 & \textbf{0.078 $\pm$ 0.01} \\
        \cline{1-7}
        \multirow[c]{2}{*}{COMPAS} & race & \isBetter 0.244 $\pm$ 0.018 & \isBetter 0.224 $\pm$ 0.02 & \isBetter 0.245 $\pm$ 0.03 & 0.266 $\pm$ 0.024 & \textbf{0.269 $\pm$ 0.026} \\
         & sex & \isBetter 0.116 $\pm$ 0.011 & \isBetter 0.111 $\pm$ 0.007 & 0.135 $\pm$ 0.011 & \isBetter 0.126 $\pm$ 0.012 & \textbf{0.137 $\pm$ 0.01} \\
        \cline{1-7}
        \multirow[c]{2}{*}{German} & age & \isBetter 0.081 $\pm$ 0.02 & \isBetter 0.095 $\pm$ 0.023 & \isBetter 0.118 $\pm$ 0.022 & \textbf{0.134 $\pm$ 0.029} & \textbf{0.134 $\pm$ 0.02} \\
         & sex & \isBetter 0.078 $\pm$ 0.014 & \isBetter 0.09 $\pm$ 0.018 & 0.115 $\pm$ 0.025 & \isWorse \textbf{0.121 $\pm$ 0.024} & 0.12 $\pm$ 0.018 \\
        \cline{1-7}
        MEPS & race & \isBetter 0.049 $\pm$ 0.003 & \isBetter 0.05 $\pm$ 0.003 & \isWorse \textbf{0.064 $\pm$ 0.004 }& \isWorse 0.063 $\pm$ 0.003 & 0.058 $\pm$ 0.005 \\
        \bottomrule
        \end{tabular}
        }
\end{table}

\begin{table*}[htb!]
\centering
    \caption{RQ2: Results comparison with different base approaches. The best approach for scenario and metric is highlighted. Winning cases are highlighted in \colorbox[HTML]{AAD5D3}{blue}, losing cases are highlighted in \colorbox[HTML]{FBC44D}{orange}.  
    }
    \label{table:basePerformance}
\begin{adjustbox}{max width=.82\textwidth}
\begin{tabular}{lll||rrrrr|rrr}
\toprule
 & \textbf{Attr}. & \textbf{Approach} & \textbf{Acc} & \textbf{Prec} & \textbf{Rec} & \textbf{F1} & \textbf{MCC} & \textbf{SPD} & \textbf{EOD} & \textbf{AOD} \\ \midrule
\multirow{10}{*}{\begin{sideways}Adult\end{sideways}} & \multirow{5}{*}{race} 
    & \tool &\textbf{0.855$\pm$0.003} & 0.829$\pm$0.003 & 0.76$\pm$0.005 & 0.784$\pm$0.005 & \textbf{0.585$\pm$0.008} & \textbf{0.073$\pm$0.005} & \textbf{0.032$\pm$0.015} & \textbf{0.024$\pm$0.01} \\
 &  & RS & \isBetter 0.844$\pm$0.005 & \isBetter 0.8$\pm$0.009 & \isWorse 0.767$\pm$0.006 & 0.781$\pm$0.006 & \isBetter 0.566$\pm$0.012 & \isBetter 0.09$\pm$0.006 & \isBetter 0.047$\pm$0.015 & \isBetter 0.04$\pm$0.01 \\
 &  & RF & \isBetter 0.841$\pm$0.002 & \isWorse 0.836$\pm$0.002 & \isWorse 0.841$\pm$0.002 & \isWorse 0.838$\pm$0.002 & \isBetter 0.559$\pm$0.005 & \isBetter 0.102$\pm$0.006 & \isBetter 0.049$\pm$0.024 & \isBetter 0.047$\pm$0.015 \\
 &  & LR & \isBetter 0.846$\pm$0.003 & \isWorse \textbf{0.839$\pm$0.003} & \isWorse \textbf{0.846$\pm$0.003} & \isWorse \textbf{0.84$\pm$0.003} & \isBetter 0.564$\pm$0.007 & \isBetter 0.099$\pm$0.007 & \isBetter 0.082$\pm$0.032 & \isBetter 0.06$\pm$0.018 \\
 &  & SVM & \isBetter 0.835$\pm$0.002 & \isBetter 0.827$\pm$0.002 & \isWorse 0.835$\pm$0.002 & \isWorse 0.827$\pm$0.002 & \isBetter 0.528$\pm$0.006 & \isBetter 0.087$\pm$0.007 & \isBetter 0.072$\pm$0.033 & \isBetter 0.052$\pm$0.017 \\  \cmidrule(lr){2-11}
 
 & \multirow{5}{*}{sex} & \tool & \textbf{0.856$\pm$0.004} & 0.828$\pm$0.004 & 0.762$\pm$0.009 & 0.786$\pm$0.007 & \textbf{0.587$\pm$0.011} & \textbf{0.164$\pm$0.005 }& 0.094$\pm$0.026 & \textbf{0.076$\pm$0.013} \\
 &  & RS & \isBetter 0.842$\pm$0.005 & \isBetter 0.797$\pm$0.008 & 0.764$\pm$0.006 & \isBetter 0.777$\pm$0.005 & \isBetter 0.559$\pm$0.01 & \isBetter 0.18$\pm$0.007 & 0.081$\pm$0.02 & 0.08$\pm$0.01 \\
 &  & RF & \isBetter 0.841$\pm$0.002 & \isWorse 0.836$\pm$0.002 & \isWorse 0.841$\pm$0.002 & \isWorse 0.838$\pm$0.002 & \isBetter 0.559$\pm$0.005 & \isBetter 0.191$\pm$0.006 & \textbf{0.08$\pm$0.022} & \isBetter 0.085$\pm$0.011 \\
 &  & LR & \isBetter 0.846$\pm$0.003 & \isWorse \textbf{0.839$\pm$0.003} & \isWorse \textbf{0.846$\pm$0.003} & \isWorse \textbf{0.84$\pm$0.003} & \isBetter 0.564$\pm$0.007 & \isBetter 0.188$\pm$0.006 & \isBetter 0.122$\pm$0.015 & \isBetter 0.102$\pm$0.008 \\
 &  & SVM & \isBetter 0.835$\pm$0.002 & 0.827$\pm$0.002 & \isWorse 0.835$\pm$0.002 & \isWorse 0.827$\pm$0.002 & \isBetter 0.528$\pm$0.006 & \isBetter 0.185$\pm$0.007 & \isBetter 0.149$\pm$0.024 & \isBetter 0.116$\pm$0.013 \\  \midrule

\multirow{5}{*}{\begin{sideways}Bank\end{sideways}} & \multirow{5}{*}{age} & \tool & \textbf{0.901$\pm$0.003} & 0.801$\pm$0.007 & 0.7$\pm$0.014 & 0.73$\pm$0.011 & 0.486$\pm$0.017 & 0.183$\pm$0.012 & 0.137$\pm$0.032 & 0.112$\pm$0.016 \\
 &  &  RS & \isBetter 0.895$\pm$0.004 & \isBetter 0.778$\pm$0.012 & 0.697$\pm$0.016 & 0.723$\pm$0.014 & \isBetter 0.465$\pm$0.021 & \isBetter 0.199$\pm$0.015 & 0.146$\pm$0.033 & \isBetter 0.127$\pm$0.016 \\
 &  & RF & 0.902$\pm$0.003 & \isWorse \textbf{0.893$\pm$0.003} & \isWorse \textbf{0.902$\pm$0.003} & \isWorse \textbf{0.896$\pm$0.003} & \isWorse \textbf{0.507$\pm$0.012} & \isBetter 0.196$\pm$0.02 & \textbf{0.126$\pm$0.047} & 0.112$\pm$0.028 \\
 &  &  LR & \isBetter 0.899$\pm$0.002 & \isWorse 0.886$\pm$0.003 & \isWorse 0.899$\pm$0.002 & \isWorse 0.888$\pm$0.003 & \isBetter 0.465$\pm$0.011 & \isBetter 0.201$\pm$0.017 & \isBetter 0.181$\pm$0.033 & \isBetter 0.142$\pm$0.02 \\
 &  &  SVM & \isBetter 0.889$\pm$0.002 & \isWorse 0.871$\pm$0.003 & \isWorse 0.889$\pm$0.002 & \isWorse 0.863$\pm$0.003 & \isBetter 0.346$\pm$0.011 & \isWorse \textbf{0.123$\pm$0.012} & \isBetter 0.159$\pm$0.029 & \textbf{0.105$\pm$0.016} \\  \midrule 
\multirow{10}{*}{\begin{sideways}COMPAS\end{sideways}} & \multirow{5}{*}{race} & \tool & 0.671$\pm$0.006 & \textbf{0.674$\pm$0.007} & 0.66$\pm$0.006 & 0.659$\pm$0.006 & 0.334$\pm$0.012 & \textbf{0.15$\pm$0.023} & \textbf{0.08$\pm$0.032} & \textbf{0.127$\pm$0.022} \\
 &  & RS & \isBetter 0.653$\pm$0.009 & \isBetter 0.65$\pm$0.009 & \isBetter 0.648$\pm$0.009 & \isBetter 0.648$\pm$0.009 & \isBetter 0.298$\pm$0.018 & 0.152$\pm$0.016 & 0.098$\pm$0.02 & 0.128$\pm$0.016 \\
 &  & RF & \isBetter 0.649$\pm$0.009 & \isBetter 0.648$\pm$0.009 & \isBetter 0.649$\pm$0.009 & \isBetter 0.648$\pm$0.009 & \isBetter 0.29$\pm$0.017 & \textbf{0.15$\pm$0.02} & 0.095$\pm$0.032 & \textbf{0.127$\pm$0.022} \\
 &  & LR & \textbf{0.673$\pm$0.01} & 0.673$\pm$0.01 & \isWorse \textbf{0.673$\pm$0.01} & \isWorse \textbf{0.669$\pm$0.01} & \textbf{0.337$\pm$0.019} & \isBetter 0.182$\pm$0.021 & \isBetter 0.109$\pm$0.03 & \isBetter 0.159$\pm$0.023 \\
 &  & SVM & \isBetter 0.658$\pm$0.01 & \isBetter 0.659$\pm$0.01 & 0.658$\pm$0.01 & \isBetter 0.651$\pm$0.011 & \isBetter 0.306$\pm$0.02 & \isBetter 0.226$\pm$0.032 & \isBetter 0.154$\pm$0.034 & \isBetter 0.207$\pm$0.033 \\ \cmidrule(lr){2-11}
 
 & \multirow{5}{*}{sex} & \tool & 0.668$\pm$0.007 &\textbf{0.673$\pm$0.007} & 0.657$\pm$0.007 & 0.654$\pm$0.009 & 0.329$\pm$0.013 & \textbf{0.088$\pm$0.021} & \textbf{0.04$\pm$0.013} & \textbf{0.057$\pm$0.02} \\
 &  & RS & \isBetter 0.658$\pm$0.009 & \isBetter 0.656$\pm$0.009 & 0.653$\pm$0.008 & 0.653$\pm$0.008 & \isBetter 0.309$\pm$0.017 & \isBetter 0.124$\pm$0.026 & \isBetter 0.067$\pm$0.021 & \isBetter 0.089$\pm$0.025 \\
 &  & RF & \isBetter 0.649$\pm$0.009 & \isBetter 0.648$\pm$0.009 & \isBetter 0.649$\pm$0.009 & \isBetter 0.648$\pm$0.009 & \isBetter 0.29$\pm$0.017 & \isBetter 0.17$\pm$0.032 & \isBetter 0.117$\pm$0.03 & \isBetter 0.137$\pm$0.034 \\
 &  & LR & \textbf{0.673$\pm$0.01} & \textbf{0.673$\pm$0.01} & \isWorse \textbf{0.673$\pm$0.01} & \isWorse \textbf{0.669$\pm$0.01} & \textbf{0.337$\pm$0.019} & \isBetter 0.265$\pm$0.031 & \isBetter  0.178$\pm$0.03 & \isBetter 0.237$\pm$0.036 \\
 &  & SVM & \isBetter 0.658$\pm$0.01 & \isBetter 0.659$\pm$0.01 & 0.658$\pm$0.01 & 0.651$\pm$0.011 & \isBetter 0.306$\pm$0.02 & \isBetter 0.279$\pm$0.027 & \isBetter 0.183$\pm$0.026 & \isBetter 0.261$\pm$0.029 \\  \midrule

\multirow{10}{*}{\begin{sideways}German\end{sideways}} & \multirow{5}{*}{age} & \tool & \textbf{0.756$\pm$0.015} & 0.722$\pm$0.02 & 0.657$\pm$0.018 & 0.668$\pm$0.019 & \textbf{0.372$\pm$0.033} & \textbf{0.058$\pm$0.028} & \textbf{0.039$\pm$0.02} & \textbf{0.041$\pm$0.017} \\
 &  & RS & \isBetter 0.744$\pm$0.018 & \isBetter 0.698$\pm$0.023 & 0.658$\pm$0.021 & 0.666$\pm$0.022 & 0.353$\pm$0.039 & \isBetter 0.096$\pm$0.045 & \isBetter 0.068$\pm$0.028 & \isBetter 0.07$\pm$0.031 \\
 &  & RF & 0.754$\pm$0.019 & \isWorse \textbf{0.742$\pm$0.02} & \isWorse \textbf{0.754$\pm$0.019} & \isWorse 0.733$\pm$0.023 & 0.364$\pm$0.043 & \isBetter 0.128$\pm$0.04 & 0.063$\pm$0.048 & \isBetter 0.089$\pm$0.049 \\
 &  & LR & 0.748$\pm$0.022 & \isWorse 0.738$\pm$0.024 & \isWorse 0.748$\pm$0.022 & \isWorse \textbf{0.737$\pm$0.024} & 0.369$\pm$0.05 & \isBetter 0.202$\pm$0.073 & \isBetter 0.159$\pm$0.074 & \isBetter 0.152$\pm$0.084 \\
 &  & SVM & 0.747$\pm$0.021 & 0.735$\pm$0.023 & \isWorse 0.747$\pm$0.021 & \isWorse 0.721$\pm$0.027 & \isBetter 0.341$\pm$0.042 & \isBetter 0.154$\pm$0.06 & \isBetter 0.094$\pm$0.059 & \isBetter 0.125$\pm$0.057 \\  \cmidrule(lr){2-11}
 
 & \multirow{5}{*}{sex} & \tool & 0.749$\pm$0.019 & 0.708$\pm$0.025 & 0.652$\pm$0.018 & 0.662$\pm$0.021 & 0.355$\pm$0.04 & \textbf{0.046$\pm$0.03} & \textbf{0.036$\pm$0.018} & \textbf{0.044$\pm$0.023} \\
 &  & RS & 0.747$\pm$0.017 & 0.704$\pm$0.021 & 0.662$\pm$0.017 & 0.67$\pm$0.018 & 0.362$\pm$0.032 & 0.057$\pm$0.033 & 0.041$\pm$0.019 & 0.05$\pm$0.03 \\
 &  & RF & \textbf{0.754$\pm$0.019} & \isWorse \textbf{0.742$\pm$0.02} & \isWorse \textbf{0.754$\pm$0.019} & \isWorse 0.733$\pm$0.023 & 0.364$\pm$0.043 & 0.061$\pm$0.045 & 0.039$\pm$0.036 & 0.055$\pm$0.041 \\
 &  & LR & 0.748$\pm$0.022 & \isWorse 0.738$\pm$0.024 & \isWorse 0.748$\pm$0.022 & \isWorse \textbf{0.737$\pm$0.024} & \textbf{0.369$\pm$0.05} & \isBetter 0.102$\pm$0.059 & 0.066$\pm$0.064 & 0.082$\pm$0.061 \\
 &  & SVM & 0.747$\pm$0.021 & \isWorse 0.735$\pm$0.023 & \isWorse 0.747$\pm$0.021 & \isWorse 0.721$\pm$0.027 & 0.341$\pm$0.042 & \isBetter 0.105$\pm$0.04 & 0.067$\pm$0.041 & \isBetter 0.091$\pm$0.052 \\  \midrule

\multirow{5}{*}{\begin{sideways}MEPS\end{sideways}} & \multirow{5}{*}{race} & \tool & 0.858$\pm$0.003 & 0.767$\pm$0.007 & 0.662$\pm$0.005 & 0.692$\pm$0.005 & 0.415$\pm$0.009 & \textbf{0.066$\pm$0.006} & \textbf{0.042$\pm$0.02} & \textbf{0.031$\pm$0.013} \\
 &  & RS & \isBetter 0.846$\pm$0.008 & \isBetter 0.741$\pm$0.019 & 0.664$\pm$0.007 & \isBetter 0.686$\pm$0.008 & \isBetter 0.395$\pm$0.018 & \isBetter 0.08$\pm$0.008 & \isBetter 0.061$\pm$0.025 & \isBetter 0.048$\pm$0.014 \\
 &  & RF & 0.859$\pm$0.003 & \isWorse 0.844$\pm$0.004 & \isWorse 0.859$\pm$0.003 & \isWorse 0.845$\pm$0.004 & \isWorse 0.43$\pm$0.013 & \isBetter 0.094$\pm$0.008 & \isBetter 0.091$\pm$0.033 & \isBetter 0.066$\pm$0.017 \\
 &  & LR & \isWorse \textbf{0.865$\pm$0.004} & \isWorse \textbf{0.851$\pm$0.005} & \isWorse \textbf{0.865$\pm$0.004} & \isWorse \textbf{0.85$\pm$0.004}& \isWorse \textbf{0.451$\pm$0.016} & \isBetter 0.113$\pm$0.008 & \isBetter 0.176$\pm$0.03 & \isBetter 0.111$\pm$0.016 \\
 &  & SVM & \isWorse 0.862$\pm$0.004 & \isWorse 0.846$\pm$0.005 & \isWorse 0.862$\pm$0.004 & \isWorse 0.843$\pm$0.005 & \isWorse 0.426$\pm$0.013 & \isBetter 0.087$\pm$0.007 & \isBetter 0.106$\pm$0.03 & \isBetter 0.07$\pm$0.015 \\ \bottomrule
\end{tabular}
\end{adjustbox}
\end{table*}

We observe how the Hypervolume returned by \tool is the highest in 5 out of 8 (62.5\%) scenarios analysed. Moreover, it is significantly better in 21 out of 32 (65.6\%) cases than other ML models. In particular, FairKNN is entirely dominated by \tool, while FairCART never wins over \tool. Hypervolumes returned by FairLR and FairSVM both win in two different scenarios. However, we also observe that the number of winning cases against those benchmarks remains higher than the number of losses.

\noindent\ding{228}\textbf{Answer to RQ1:} RFs (\tool) work best for 62.5\% of the cases, followed by FairLR and FairSVM classifiers.

\subsection{Answer to RQ2 -- \tool vs Base Strategies}
\label{sec:Results:RQ2}

\begin{figure*}[tb]
\centering
  \includegraphics[width=.7\linewidth]{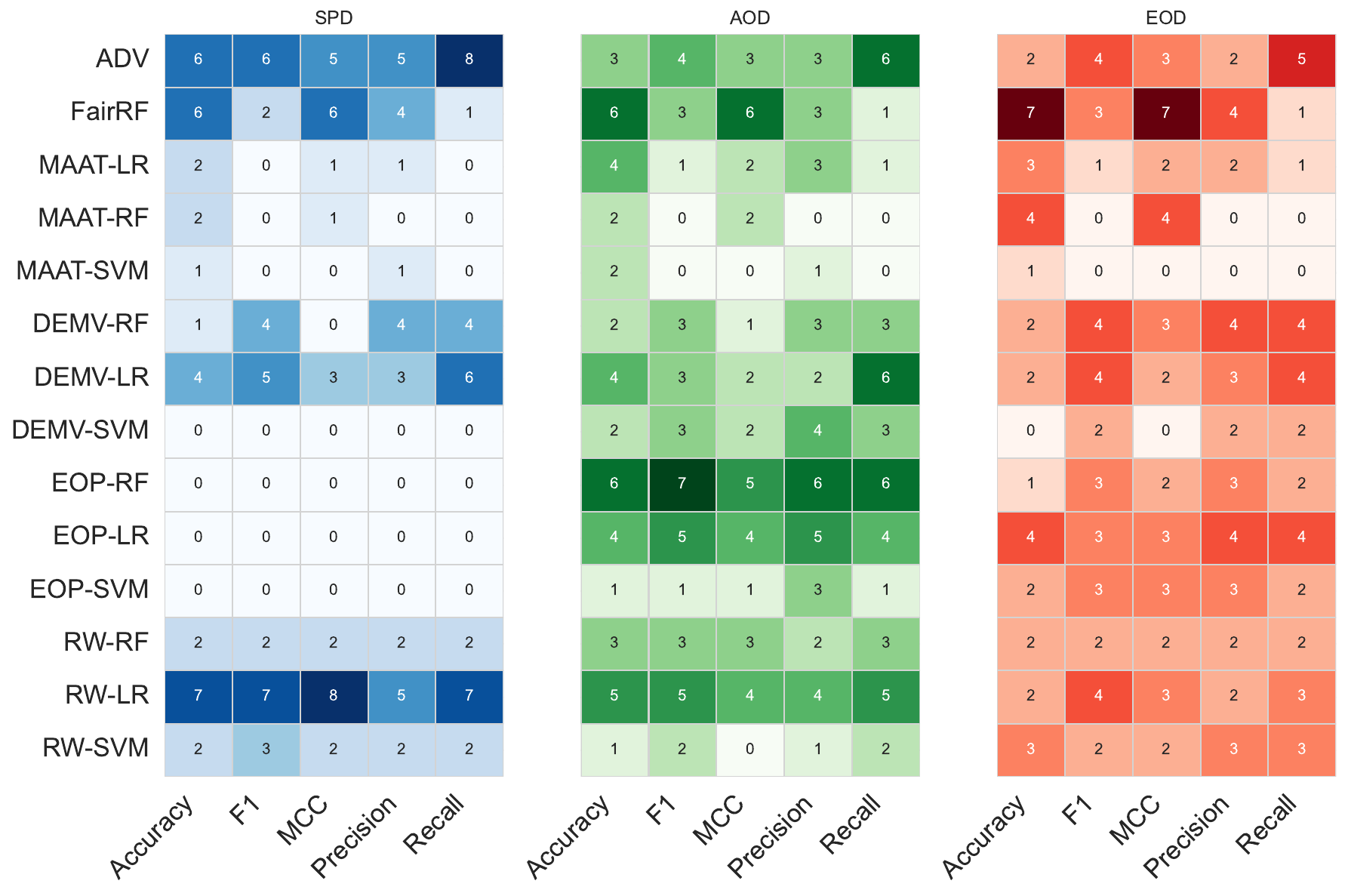}
    \caption{RQ3: Pareto optimality of each approach considering each pair of effectiveness and fairness metrics. 
    }
  \label{fig:heatmapPareto}
\end{figure*}

Table \ref{table:basePerformance} reports the fairness and performance scores obtained by \tool and the base strategies for each scenario analysed. To improve readability, we have omitted those from the KNN and CART models, since their effectiveness score was always very low, indicating their lower effectiveness compared to other models. However, those are available in our replication package \replpackage. As done for \textbf{RQ1}, for each analysed scenario (i.e., dataset plus sensitive attribute), we highlight in \textbf{bold} the best metrics, in \colorbox{lightblue}{blue} cases where the score obtained by \tool is significantly better, and in \colorbox{orange}{orange} cases where the score obtained by \tool is worse.

From the table, we observe that \tool achieves a significantly better fairness than base strategies for each of the three metrics. In particular, \tool provides the best SPD score in 7 out of 8 (87.5\%) scenarios, the best EOD score in 6 out of 8 scenarios (75\%), and the best AOD score also in 7 out of 8 (87.5\%) cases. Additionally, the scores obtained by \tool are significantly larger in the majority of cases (84.7\% for SPD, 65.6\% for EOD, and 75\% for AOD). Particularly, we observe just one scenario (i.e., Bank -- age) in which the SPD score obtained by \tool is worse than the one obtained by the SVM model. However, the SVM model in the same scenario provides a significant drop in Accuracy and MCC compared to \tool.

Finally, the fairness improvement obtained by \tool does not imply a significant degradation in prediction effectiveness. In fact, \tool is able to improve the Accuracy of base strategies in a significant manner in 19 out of 32 (59.3\%) cases, with only one scenario (i.e., MEPS -- race) in which Accuracy is dominated by LR and SVM models. We observe an overall decrease in Precision, Recall, and F1 Score, which may be expected in the case of fairness improvement \cite{hort_bias_2023}. However, when considering MCC, which is a more comprehensive metric that considers all the squares of a confusion matrix \cite{moussa2022use,Menzies2007Response}, we observe how \tool provides a score that is significantly better than the benchmarks in 56.2\% of the cases and significantly worse in only 12.5\% of the cases.

\noindent\ding{228}\textbf{Answer to RQ2:} In terms of fairness improvement, \tool overcomes base strategies in a statistically significant manner in 84.7\% of cases concerning SPD, 65.6\% of cases concerning EOD, and 75\% of cases concerning EOD. At the same time, \tool provides significant improvements in terms of Accuracy and the most comprehensive MCC score.

\begin{figure}[tb]
    \centering
    \includegraphics[width=.5\textwidth]{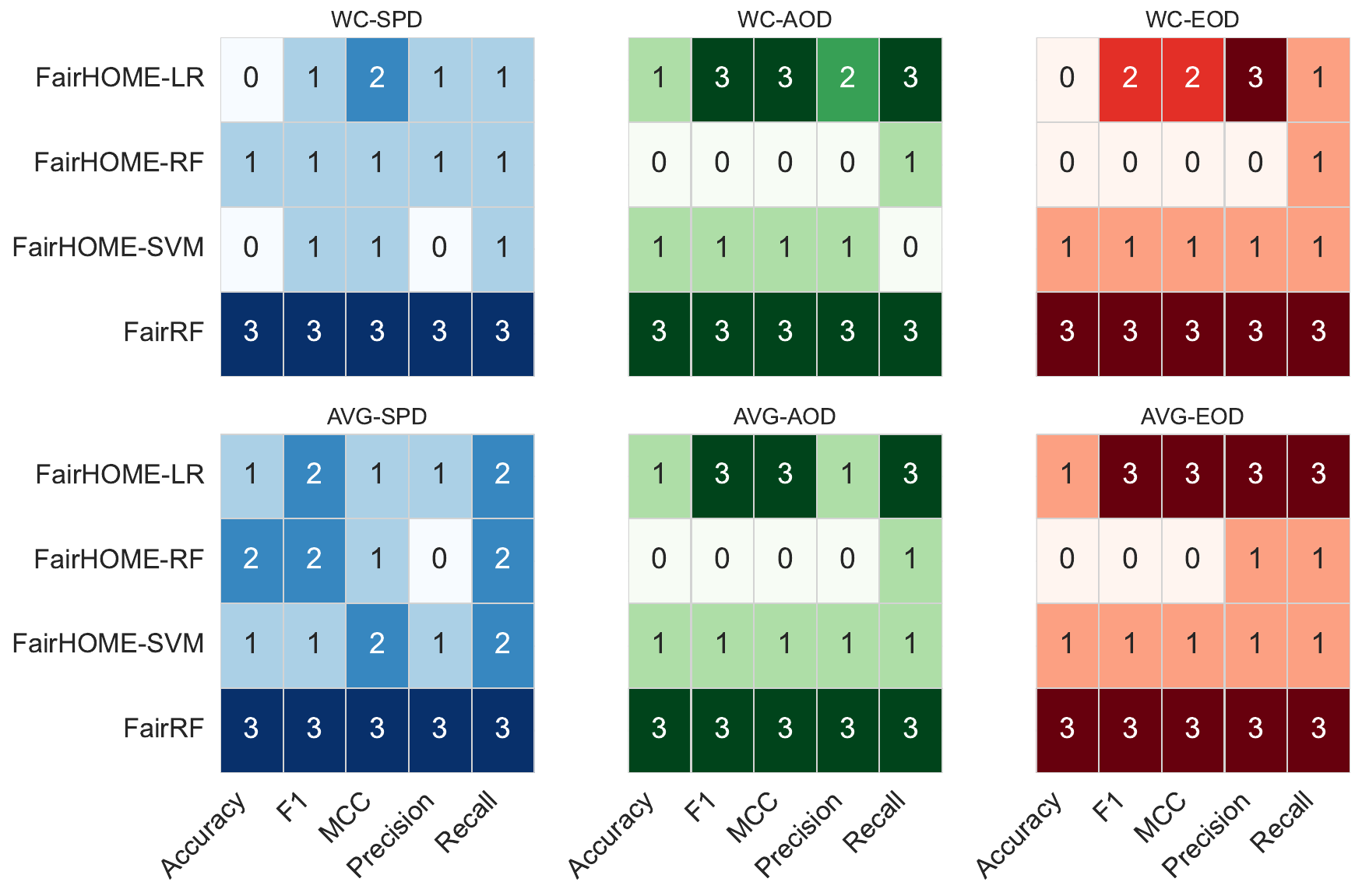}
    \caption{RQ4: Pareto optimality of each approach considering each pair of effectiveness and intersectional fairness metrics.}
    \label{fig:intersectional}
\end{figure}

\subsection{Answer to RQ3 -- \tool vs State-Of-The-Art Bias Mitigation Methods}
\label{sec:Results:RQ3}

Figure \ref{fig:heatmapPareto} reports the Pareto optimality distribution of \tool against SOTA approaches. In computing the Pareto optimality, we consider the combination of each fairness and effectiveness metric and aggregate the results for all scenarios analysed.\footnote{Scenario-specific results are available in our replication package \replpackage.} Note that when evaluating pre-processing and post-processing methods, we consider LR, RF and SVM as base classifiers, since they have been shown to be the most effective approaches from \textbf{RQ2}. 

From the figure, we observe how RW for LR models is the best approach when considering the combination of SPD and effectiveness metrics. On the contrary, EOP for RF models is the optimal approach, considering the combination of AOD and effectiveness metrics. Finally, no approach emerges as a clear winner when examining the combination of EOD and effectiveness metrics. However, we observe that \tool consistently provides a high number of Pareto optimal solutions under the combination of Accuracy, MCC, and all fairness metrics. These results are in line with those observed for \textbf{RQ2} and suggest that, when looking at all the squares of a confusion matrix (i.e., TP, FP, TN, and FN), \tool overcomes SOTA approaches under all fairness definitions considered.

\noindent\ding{228}\textbf{Answer to RQ3}:
\tool overcomes SOTA bias mitigation approaches under all fairness definitions when looking at both Accuracy and MCC scores.

\subsection{Answer to RQ4 -- \tool vs State-Of-The-Art Intersectional Bias Mitigation Method}

Figure \ref{fig:intersectional} reports the number of Pareto optimal solutions reported by each approach for each effectiveness and intersectional fairness metric combination aggregated for all scenarios. As done for \textbf{RQ3}, we employ LR, RF, and SVM as base classifiers for FairHOME.

From the plot, we can see that \tool outperforms all versions of FairHOME across all combinations of fairness and effectiveness metrics. Specifically, \tool consistently delivers a stable number of Pareto optimal solutions regardless of the metric combinations used. In contrast, the various versions of FairHOME show an inconsistent number of Pareto optimal solutions. Notably, we did not find any combination of metrics where a version of FairHOME yields a higher number of Pareto-optimal solutions than \tool.
This result highlights the efficacy of \tool in the intersectional fairness improvement, overcoming the SOTA FairHOME approach. 

\noindent\ding{228}\textbf{Answer to RQ4}: \tool overcomes the SOTA FairHOME approach under all combinations of effectiveness and intersectional fairness metrics.

\subsection{Discussion}
\label{sec:advantages}

From our evaluation, it can be seen how \tool is a valuable solution for bias mitigation in both single and intersectional 
scenarios.

First, we observed that optimising for SPD alone yields significant results under other fairness definitions as well. Specifically, considering the average scores among all the scenarios (dataset -- sensitive attribute combinations) analysed, \tool provides a maximum improvement in SPD of 63.42\% against LR and a minimum improvement of 18.15\% against RF. Moreover, \tool achieves a maximum improvement in EOD of 114.93\% against LR and a minimum of 10.33\% against CART. Finally, it achieves a maximum improvement in AOD of 104.37\%, still against LR and a minimum of 22.47\% against CART. Moreover, as observed in \textbf{RQ2}, the mutation in the input dataset provides a significant contribution to bias mitigation, since \tool always achieved significantly better fairness scores against the base RF classifier. Finally, \tool behaves consistently in bias mitigation against SOTA approaches for single sensitive variable bias mitigation, while significantly overcoming the SOTA for intersectional bias mitigation.

Regarding the effectiveness results, we noted that \tool generally yields lower scores in Precision, Recall, and F1 Score compared to the base classifier. This outcome may be expected, as improvements in fairness often lead to a reduction in positive outcomes relative to base classifiers \cite{10.1145/3729350, hort_bias_2023}. 
However, when examining the distribution of Pareto optimal solutions shown in Figure \ref{fig:heatmapPareto}, we see that \tool exhibits lower Pareto optimal results in terms of Recall compared to SOTA approaches. This observation may suggest that \tool has a reduced capacity for delivering positive outcomes. Nevertheless, when it does succeed, those outcomes tend to be accurate; this is evidenced by a consistent number of Pareto optimal solutions in terms of Accuracy, Precision, and MCC compared to SOTA methods. 
This behaviour of \tool may be explained by the choice of Accuracy as a fitness function to guide the search. However, an advantage of \tool compared to other approaches is its flexibility and adaptability to the stakeholders' needs. In fact, in cases where it may be more important to maximise the number of true positives (e.g., safety-critical software systems like autonomous driving \cite{li_bias_2025} or health devices \cite{canali_challenges_2022}), data scientists and engineers can extend \tool by employing different fitness functions to guide the search. Additionally, they may also choose to give different weights (i.e., priorities) to either effectiveness or fairness fitness functions. This flexibility of \tool makes it a novel contribution to the topic of bias mitigation, allowing stakeholders to adapt this task based on their specific needs. Finally, we noticed that, even with this setting (i.e., optimising for Accuracy and SPD), at least one solution from \tool has always been present in the Pareto front for any combination of effectiveness and fairness metrics, making it a valuable option for practitioners to choose from.

\section{Threats to Validity}
\label{sec:Threats}

\textit{External Validity:} The reported results might not generalise to real-world scenarios. To mitigate this, we considered several classification models and popular datasets to show that \tool is effective in diverse scenarios.

\noindent\textit{Internal Validity:} We employ Accuracy and SPD as fitness functions to guide the search. While these metrics may partially represent fairness and effectiveness of a model, we show how the results obtained using this setting provide consistent results even under different definitions. We also discuss how \tool can be easily extended by data scientists and engineers to employ different fitness functions based on their needs. 

\noindent\textit{Construct Validity:} The stochastic nature of ML models and the bias mitigation methods applied to improve their fairness raise concerns about the representativeness of measured results (i.e., fairness and effectiveness).
To miitigate this, we performed experiments 20 times and reported the mean performance in addition to the standard deviation.
Moreover, we conducted statistical tests (i.e., hypothesis testing and effect size) to evaluate the performance of \tool and compare it against baselines as well as state-of-the-art bias mitigation methods. 
Also, implementations of existing algorithms and datasets are obtained from AIF360 or respective GitHub repositories, thus minimising the risk of faults in implementations.

\section{Conclusions and Future Work}
\label{sec:Conclusions}

In this paper, we proposed \tool, a MOEA that searches for RF hyperparameters and data mutations to optimise fairness and effectiveness. We performed an extensive evaluation of \tool against 26 baselines in 11 different scenarios, highlighting the advantages of \tool over existing approaches.

Future work can explore the adoption of \tool to improve fairness in less-explored tasks like multi-class classification \cite{daloisio_debiaser_2023}. Additionally, \tool can be extended to explore a larger hyperparameter space or different data mutation strategies. Finally, approaches to better guide stakeholders in understanding the Pareto solutions returned by \tool can be investigated.

\begin{acks}

    G. d'Aloisio is partially funded by "SoBigData.it - Strengthening the Italian RI for Social Mining and Big Data Analytics" - Prot. IR0000013 - Avviso n. 3264 del 28/12/2021. M. Hort is supported by the the European Union through Horizon Europe Marie Sk\l{}odowska-Curie Actions (\#101151798).
\end{acks}

\balance

\bibliographystyle{ACM-Reference-Format}
\bibliography{bib}

@inproceedings{gong_2025,
      title={GA4GC: Greener Agent for Greener Code via Multi-Objective Configuration Optimization}, 
      author={Jingzhi Gong and Yixin Bian and Luis de la Cal and Giovanni Pinna and Anisha Uteem and David Williams and Mar Zamorano and Karine Even-Mendoza and W. B. Langdon and Hector Menendez and Federica Sarro},
      year={2025},
      eprint={2510.04135},
      url={https://arxiv.org/abs/2510.04135},
booktitle={International Symposioum on Search Based Software Engineering (SSBSE)}
}

@inproceedings{moussa2022use,
  title={On the Use of Evaluation Measures for Defect Prediction Studies},
  author={Moussa, Rebecca and Sarro, Federica},
  year={2022},
  organization={ACM},
booktitle={2022 ACM SIGSOFT International Symposium on Software Testing and Analysis (ISSTA)}
}

@ARTICLE{Menzies2007Response,
  author={Menzies, Tim and Dekhtyar, Alex and Distefano, Justin and Greenwald, Jeremy},
  journal={IEEE Transactions on Software Engineering}, 
  title={Problems with Precision: A Response to "Comments on 'Data Mining Static Code Attributes to Learn Defect Predictors'"}, 
  year={2007},
  volume={33},
  number={9},
  pages={637-640},
  doi={10.1109/TSE.2007.70721}}

@inproceedings{Sarro19Keynote,
  author    = {Federica Sarro},
  title     = {Search-Based Predictive Modelling for Software Engineering: How Far
               Have We Gone?},
  booktitle = {Search-Based Software Engineering - 11th International Symposium,
               {SSBSE} 2019, Tallinn, Estonia, August 31 - September 1, 2019, Proceedings},
  series    = {Lecture Notes in Computer Science},
  volume    = {11664},
  pages     = {3--7},
  publisher = {Springer},
  year      = {2019},
  url       = {https://doi.org/10.1007/978-3-030-27455-9\_1},
  doi       = {10.1007/978-3-030-27455-9\_1},
}

@Article{Arcuri2014,
  author  = {Arcuri, Andrea and Briand, Lionel},
  title   = {{A Hitchhiker's guide to statistical tests for assessing randomized algorithms in software engineering}},
  doi     = {10.1002/stvr.1486},
  issn    = {0960-0833},
  number  = {3},
  pages   = {219--250},
  volume  = {24},
  journal = {Software Testing, Verification and Reliability},
  month   = may,
  year    = {2014},
}

@Article{Harman2011,
  author          = {Harman, Mark and McMinn, Phil and {De Souza}, Jerffeson Teixeira and Yoo, Shin},
  title           = {{Search based software engineering: Techniques, taxonomy, tutorial}},
  doi             = {10.1007/978-3-642-25231-0_1},
  issn            = {0302-9743},
  pages           = {1--59},
  volume          = {7007 LNCS},
  isbn            = {9783642252303},
  journal         = {Lecture Notes in Computer Science (including subseries Lecture Notes in Artificial Intelligence and Lecture Notes in Bioinformatics)},
  year            = {2011},
}

@InCollection{Han_2021,
  author    = {Kate Han and Tien Pham and Trung Hieu Vu and Truong Dang and John McCall and Tien Thanh Nguyen},
  booktitle = {Intelligent Information and Database Systems},
  title     = {{VEGAS}: A Variable Length-Based Genetic Algorithm for Ensemble Selection in Deep Ensemble Learning},
  doi       = {10.1007/978-3-030-73280-6_14},
  pages     = {168--180},
  publisher = {Springer International Publishing},
  year      = {2021},
}

@Article{Fletcher_2020,
  author    = {Sam Fletcher and Brijesh Verma and Mengjie Zhang},
  title     = {A non-specialized ensemble classifier using multi-objective optimization},
  doi       = {10.1016/j.neucom.2020.05.029},
  pages     = {93--102},
  volume    = {409},
  journal   = {Neurocomputing},
  month     = oct,
  publisher = {Elsevier {BV}},
  year      = {2020},
}

@inproceedings{Chen_2018,
author = {Chen, Boyuan and Wu, Harvey and Mo, Warren and Chattopadhyay, Ishanu and Lipson, Hod},
title = {{Autostacker: A Compositional Evolutionary Learning System}},
year = {2018},
doi = {10.1145/3205455.3205586},
booktitle = {Proceedings of the Genetic and Evolutionary Computation Conference},
pages = {402–-409},
numpages = {8},
location = {Kyoto, Japan},
series = {GECCO'18}
}

@InProceedings{Dos_Santos_2006,
  author    = {E.M. Dos Santos and R. Sabourin and P. Maupin},
  booktitle = {The 2006 {IEEE} International Joint Conference on Neural Network Proceedings},
  title     = {Single and Multi-Objective Genetic Algorithms for the Selection of Ensemble of Classifiers},
  doi       = {10.1109/ijcnn.2006.247267},
  publisher = {{IEEE}},
  year      = {2006},
}

@article{hort_bias_2023,
    title = {Bias {Mitigation} for {Machine} {Learning} {Classifiers}: {A} {Comprehensive} {Survey}},
    issn = {2832-0565},
    shorttitle = {Bias {Mitigation} for {Machine} {Learning} {Classifiers}},
    url = {https://dl.acm.org/doi/10.1145/3631326},
    doi = {10.1145/3631326},
    language = {en},
    urldate = {2023-12-13},
    journal = {ACM Journal on Responsible Computing},
    author = {Hort, Max and Chen, Zhenpeng and Zhang, Jie M. and Harman, Mark and Sarro, Federica},
    month = nov,
    year = {2023},
    pages = {3631326},
}

@misc{chen_diversity_2024,
    title = {Diversity {Drives} {Fairness}: {Ensemble} of {Higher} {Order} {Mutants} for {Intersectional} {Fairness} of {Machine} {Learning} {Software}},
    shorttitle = {Diversity {Drives} {Fairness}},
    url = {http://arxiv.org/abs/2412.08167},
    doi = {10.48550/arXiv.2412.08167},
    urldate = {2024-12-16},
    publisher = {arXiv},
    author = {Chen, Zhenpeng and Li, Xinyue and Zhang, Jie M. and Sarro, Federica and Liu, Yang},
    month = dec,
    year = {2024},
    note = {arXiv:2412.08167 [cs]},
}

@inproceedings{sarro_multi-objective_2016,
	address = {New York, NY, USA},
	series = {{ICSE} '16},
	title = {Multi-objective software effort estimation},
	isbn = {978-1-4503-3900-1},
	url = {https://dl.acm.org/doi/10.1145/2884781.2884830},
	doi = {10.1145/2884781.2884830},
	urldate = {2024-01-14},
	booktitle = {Proceedings of the 38th {International} {Conference} on {Software} {Engineering}},
	publisher = {Association for Computing Machinery},
	author = {Sarro, Federica and Petrozziello, Alessio and Harman, Mark},
	month = may,
	year = {2016},
	pages = {619--630},
}

@ARTICLE{8637212,
  author={Ishibuchi, Hisao and Imada, Ryo and Setoguchi, Yu and Nojima, Yusuke},
  journal={Evolutionary Computation}, 
  title={How to Specify a Reference Point in Hypervolume Calculation for Fair Performance Comparison}, 
  year={2018},
  volume={26},
  number={3},
  pages={411-440},
  doi={10.1162/evco_a_00226}}

@article{10.1145/3729350,
author = {Chen, Zhenpeng and Li, Xinyue and Zhang, Jie M. and Sun, Weisong and Xiao, Ying and Li, Tianlin and Lou, Yiling and Liu, Yang},
title = {Software Fairness Dilemma: Is Bias Mitigation a Zero-Sum Game?},
year = {2025},
issue_date = {July 2025},
publisher = {Association for Computing Machinery},
address = {New York, NY, USA},
volume = {2},
number = {FSE},
url = {https://doi.org/10.1145/3729350},
doi = {10.1145/3729350},
journal = {Proc. ACM Softw. Eng.},
month = jun,
articleno = {FSE080},
numpages = {22}
}

@article{10.1145/3453474,
author = {Guerreiro, Andreia P. and Fonseca, Carlos M. and Paquete, Lu\'{\i}s},
title = {The Hypervolume Indicator: Computational Problems and Algorithms},
year = {2021},
issue_date = {July 2022},
publisher = {Association for Computing Machinery},
address = {New York, NY, USA},
volume = {54},
number = {6},
issn = {0360-0300},
url = {https://doi.org/10.1145/3453474},
doi = {10.1145/3453474},
journal = {ACM Comput. Surv.},
month = jul,
articleno = {119},
numpages = {42}
}

@incollection{wilcoxon1992individual,
  title={Individual comparisons by ranking methods},
  author={Wilcoxon, Frank},
  booktitle={Breakthroughs in statistics: Methodology and distribution},
  pages={196--202},
  year={1992},
  publisher={Springer}
}

@inproceedings{moussa_meg_2022,
	address = {Helsinki Finland},
	title = {{MEG}: {Multi}-objective {Ensemble} {Generation} for {Software} {Defect} {Prediction}},
	isbn = {978-1-4503-9427-7},
	shorttitle = {{MEG}},
	url = {https://dl.acm.org/doi/10.1145/3544902.3546255},
	doi = {10.1145/3544902.3546255},
	language = {en},
	urldate = {2024-03-12},
	booktitle = {Proceedings of the 16th {ACM} / {IEEE} {International} {Symposium} on {Empirical} {Software} {Engineering} and {Measurement}},
	publisher = {ACM},
	author = {Moussa, Rebecca and Guizzo, Giovani and Sarro, Federica},
	month = sep,
	year = {2022},
	pages = {159--170},
}

@article{daloisio_debiaser_2023,
    title = {Debiaser for {Multiple} {Variables} to enhance fairness in classification tasks},
    volume = {60},
    copyright = {All rights reserved},
    issn = {0306-4573},
    url = {https://www.sciencedirect.com/science/article/pii/S0306457322003272},
    doi = {10.1016/j.ipm.2022.103226},
    language = {en},
    number = {2},
    urldate = {2022-12-22},
    journal = {Information Processing \& Management},
    author = {d’Aloisio, Giordano and D’Angelo, Andrea and Di Marco, Antinisca and Stilo, Giovanni},
    month = mar,
    year = {2023},
    pages = {103226},
}

@article{deb2002nsga2,
    author = {{Deb, K.}},
    title = {{A fast and elitist multiobjective genetic algorithm}: NSGA-II},
    journal = {ieeexplore},
    year = {2002}
}

@inproceedings{zhang2018mitigating,
  title={Mitigating unwanted biases with adversarial learning},
  author={Zhang, Brian Hu and Lemoine, Blake and Mitchell, Margaret},
  booktitle={Proceedings of the 2018 AAAI/ACM Conference on AI, Ethics, and Society},
  pages={335--340},
  year={2018}
}

@inproceedings{chakraborty2021bias,
  title={Bias in machine learning software: Why? how? what to do?},
  author={Chakraborty, Joymallya and Majumder, Suvodeep and Menzies, Tim},
  booktitle={Proceedings of the 29th ACM joint meeting on European software engineering conference and symposium on the foundations of software engineering},
  pages={429--440},
  year={2021}
}

@article{DEAP_JMLR2012,
  author    = {F\'elix-Antoine Fortin and Fran\c{c}ois-Michel De Rainville and Marc-Andr\'e Gardner and Marc Parizeau and Christian Gagn\'e},
  title     = {{DEAP}: Evolutionary Algorithms Made Easy},
  journal   = {Journal of Machine Learning Research},
  volume    = {13},
  pages     = {2171--2175},
  year      = {2012},
  month     = {7}
}

@inproceedings{hardt2016equality,
  title={Equality of opportunity in supervised learning},
  author={Hardt, Moritz and Price, Eric and Srebro, Nati},
  booktitle={Advances in neural information processing systems},
  pages={3315--3323},
  year={2016}
}

@article{kamiran2012data,
  title={Data preprocessing techniques for classification without discrimination},
  author={Kamiran, Faisal and Calders, Toon},
  journal={Knowledge and information systems},
  volume={33},
  number={1},
  pages={1--33},
  year={2012},
  publisher={Springer}
}

@article{zhang2020machine,
  title={Machine learning testing: Survey, landscapes and horizons},
  author={Zhang, Jie M and Harman, Mark and Ma, Lei and Liu, Yang},
  journal={IEEE Transactions on Software Engineering},
  volume={48},
  number={1},
  pages={1--36},
  year={2020},
  publisher={IEEE}
}

@inproceedings{beutel2019putting,
  title={Putting fairness principles into practice: Challenges, metrics, and improvements},
  author={Beutel, Alex and Chen, Jilin and Doshi, Tulsee and Qian, Hai and Woodruff, Allison and Luu, Christine and Kreitmann, Pierre and Bischof, Jonathan and Chi, Ed H},
  booktitle={Proceedings of the 2019 AAAI/ACM Conference on AI, Ethics, and Society},
  pages={453--459},
  year={2019}
}

@MISC {adultdata,
  author       = {Becker, Barry and Kohavi, Ronny},
  title        = {{Adult}},
  year         = {1996},
  howpublished = {UCI Machine Learning Repository},
  note         = {{DOI}: https://doi.org/10.24432/C5XW20}
}

@misc{replpackage,
    author = {d'Aloisio, Giordano and Hort, Max and Moussa, Rebecca and Sarro, Federica},
    title = {Fair RF Replication Package},
    url = {https://github.com/SOLAR%2Dgroup/FairRF},
    year = 2026
}

@software{d_aloisio_2025_17879088,
  author       = {d'Aloisio, Giordano and
                  Hort, Max and
                  Moussa, Rebecca and
                  Sarro, Federica},
  title        = {SOLAR-group/FairRF: FairRF},
  month        = dec,
  year         = 2025,
  publisher    = {Zenodo},
  version      = {1.0},
  doi          = {10.5281/zenodo.17879088},
  url          = {https://doi.org/10.5281/zenodo.17879088},
}

@article{moro2014data,
  title={A data-driven approach to predict the success of bank telemarketing},
  author={Moro, S{\'e}rgio and Cortez, Paulo and Rita, Paulo},
  journal={Decision Support Systems},
  volume={62},
  pages={22--31},
  year={2014},
  publisher={Elsevier}
}

@MISC {compas,
    author       = "propublica",
    title        = "data for the propublica story ‘machine bias’",
    howpublished = "\url{https://github.com/propublica/compas-analysis/}",
    year = 2016
}

@MISC {germandata,
    author       = "Dr. Hans Hofmann",
    title        = "Statlog (german credit data) data set",
    howpublished = "\url{http://archive.ics.uci.edu/ml/datasets/statlog+(german+credit+data)}",
    year = 2016
}

@inproceedings{daloisio_sustaindiffusion_2025,
    title = {{SustainDiffusion}: {Optimising} the {Social} and {Environmental} {Sustainability} of {Stable} {Diffusion} {Models}},
    url = {https://arxiv.org/abs/2507.15663},
    author = {d'Aloisio, Giordano and Fadahunsi, Tosin and Choy, Jay and Moussa, Rebecca and Sarro, Federica},
    booktitle={48th IEEE/ACM International Conference on Software Engineering, ICSE 2026},
    year = {2026}
}

@inproceedings{friedler2019comparative,
  title={A comparative study of fairness-enhancing interventions in machine learning},
  author={Friedler, Sorelle A and Scheidegger, Carlos and Venkatasubramanian, Suresh and Choudhary, Sonam and Hamilton, Evan P and Roth, Derek},
  booktitle={Proceedings of the Conference on Fairness, Accountability, and Transparency},
  pages={329--338},
  year={2019},
  organization={ACM}
}

@inproceedings{calders2009building,
  title={Building classifiers with independency constraints},
  author={Calders, Toon and Kamiran, Faisal and Pechenizkiy, Mykola},
  booktitle={2009 IEEE International Conference on Data Mining Workshops},
  pages={13--18},
  year={2009},
  organization={IEEE}
}

@article{tizpaz2022fairness,
  title={Fairness-aware Configuration of Machine Learning Libraries},
  author={Tizpaz-Niari, Saeid and Kumar, Ashish and Tan, Gang and Trivedi, Ashutosh},
  journal={arXiv preprint arXiv:2202.06196},
  year={2022}
}

@article{savani2020intra,
  title={Intra-processing methods for debiasing neural networks},
  author={Savani, Yash and White, Colin and Govindarajulu, Naveen Sundar},
  journal={Advances in Neural Information Processing Systems},
  volume={33},
  pages={2798--2810},
  year={2020}
}

@inproceedings{kamiran2010discrimination,
  title={Discrimination aware decision tree learning},
  author={Kamiran, Faisal and Calders, Toon and Pechenizkiy, Mykola},
  booktitle={2010 IEEE International Conference on Data Mining},
  pages={869--874},
  year={2010},
  organization={IEEE}
}

@inproceedings{speicher2018unified,
  title={A unified approach to quantifying algorithmic unfairness: Measuring individual \&group unfairness via inequality indices},
  author={Speicher, Till and Heidari, Hoda and Grgic-Hlaca, Nina and Gummadi, Krishna P and Singla, Adish and Weller, Adrian and Zafar, Muhammad Bilal},
  booktitle={Proceedings of the 24th ACM SIGKDD International Conference on Knowledge Discovery \& Data Mining},
  pages={2239--2248},
  year={2018}
}

@inproceedings{dwork2018decoupled,
  title={Decoupled classifiers for group-fair and efficient machine learning},
  author={Dwork, Cynthia and Immorlica, Nicole and Kalai, Adam Tauman and Leiserson, Max},
  booktitle={Conference on fairness, accountability and transparency},
  pages={119--133},
  year={2018},
  organization={PMLR}
}

@article{pleiss2017fairness,
  title={On fairness and calibration},
  author={Pleiss, Geoff and Raghavan, Manish and Wu, Felix and Kleinberg, Jon and Weinberger, Kilian Q},
  journal={Advances in neural information processing systems},
  volume={30},
  year={2017}
}

@inproceedings{dwork2012fairness,
  title={Fairness through awareness},
  author={Dwork, Cynthia and Hardt, Moritz and Pitassi, Toniann and Reingold, Omer and Zemel, Richard},
  booktitle={Proceedings of the 3rd innovations in theoretical computer science conference},
  pages={214--226},
  year={2012}
}

@inproceedings{chen2022maat,
  title={MAAT: a novel ensemble approach to addressing fairness and performance bugs for machine learning software},
  author={Chen, Zhenpeng and Zhang, Jie M and Sarro, Federica and Harman, Mark},
  booktitle={Proceedings of the 30th ACM joint european software engineering conference and symposium on the foundations of software engineering},
  pages={1122--1134},
  year={2022}
}

@book{breiman2017classification,
  title={Classification and regression trees},
  author={Breiman, Leo and Friedman, Jerome and Olshen, Richard A and Stone, Charles J},
  year={2017},
  publisher={Chapman and Hall/CRC}
}

@article{hort_search-based_2024,
    title = {Search-based {Automatic} {Repair} for {Fairness} and {Accuracy} in {Decision}-making {Software}},
    volume = {29},
    issn = {1573-7616},
    url = {https://doi.org/10.1007/s10664-023-10419-3},
    doi = {10.1007/s10664-023-10419-3},
    language = {en},
    number = {1},
    urldate = {2024-01-14},
    journal = {Empirical Software Engineering},
    author = {Hort, Max and Zhang, Jie M. and Sarro, Federica and Harman, Mark},
    month = jan,
    year = {2024},
    pages = {36},
}

@article{rosenfield_coefficient_1986,
    title = {A coefficient of agreement as a measure of thematic classification accuracy.},
    volume = {52},
    url = {http://pubs.er.usgs.gov/publication/70014667},
    number = {2},
    journal = {Photogrammetric Engineering and Remote Sensing},
    author = {Rosenfield, G.H. and Fitzpatrick-Lins, K.},
    year = {1986},
    pages = {223--227},
}

@article{pedregosa_scikit-learn_2011,
    title = {Scikit-learn: {Machine} {Learning} in {Python}},
    volume = {12},
    journal = {Journal of Machine Learning Research},
    author = {Pedregosa, F. and Varoquaux, G. and Gramfort, A. and Michel, V. and Thirion, B. and Grisel, O. and Blondel, M. and Prettenhofer, P. and Weiss, R. and Dubourg, V. and Vanderplas, J. and Passos, A. and Cournapeau, D. and Brucher, M. and Perrot, M. and Duchesnay, E.},
    year = {2011},
    pages = {2825--2830},
}

@article{bellamy_ai_2019,
    title = {{AI} {Fairness} 360: {An} extensible toolkit for detecting and mitigating algorithmic bias},
    volume = {63},
    issn = {0018-8646},
    shorttitle = {{AI} {Fairness} 360},
    doi = {10.1147/JRD.2019.2942287},
    number = {4/5},
    journal = {IBM Journal of Research and Development},
    author = {Bellamy, R. K. E. and Dey, K. and Hind, M. and Hoffman, S. C. and Houde, S. and Kannan, K. and Lohia, P. and Martino, J. and Mehta, S. and Mojsilović, A. and Nagar, S. and Ramamurthy, K. Natesan and Richards, J. and Saha, D. and Sattigeri, P. and Singh, M. and Varshney, K. R. and Zhang, Y.},
    month = jul,
    year = {2019},
    note = {Conference Name: IBM Journal of Research and Development},
    pages = {4:1--4:15},
}

@article{buckland_relationship_1994,
    title = {The relationship between recall and precision},
    volume = {45},
    number = {1},
    journal = {Journal of the American society for information science},
    author = {Buckland, Michael and Gey, Fredric},
    year = {1994},
    note = {Publisher: Wiley Online Library},
    pages = {12--19},
}

@article{mehrabi_survey_2021,
    title = {A {Survey} on {Bias} and {Fairness} in {Machine} {Learning}},
    volume = {54},
    issn = {0360-0300, 1557-7341},
    doi = {10.1145/3457607},
    language = {en},
    number = {6},
    urldate = {2021-11-11},
    journal = {ACM Computing Surveys},
    author = {Mehrabi, Ninareh and Morstatter, Fred and Saxena, Nripsuta and Lerman, Kristina and Galstyan, Aram},
    month = jul,
    year = {2021},
    pages = {1--35},
}

@article{caton_fairness_2023,
    title = {Fairness in {Machine} {Learning}: {A} {Survey}},
    issn = {0360-0300},
    shorttitle = {Fairness in {Machine} {Learning}},
    url = {https://dl.acm.org/doi/10.1145/3616865},
    doi = {10.1145/3616865},
    urldate = {2023-12-04},
    journal = {ACM Computing Surveys},
    author = {Caton, Simon and Haas, Christian},
    year = {2023},
    note = {Just Accepted},
}

@misc{chen_fairness_2024,
    type = {Proceedings paper},
    title = {Fairness {Improvement} with {Multiple} {Protected} {Attributes}: {How} {Far} {Are} {We}?},
    copyright = {open},
    shorttitle = {Fairness {Improvement} with {Multiple} {Protected} {Attributes}},
    url = {https://www.computer.org/csdl/proceedings/icse/2024/1RLIVDkr2xO},
    language = {eng},
    urldate = {2024-02-27},
    journal = {In:  Proceedings of the 2024 IEEE/ACM 46th International Conference on Software Engineering (ICSE).    IEEE/ACM: Lisbon, Portugal. (2024)    (In press).},
    author = {Chen, Zhenpeng and Zhang, Jie M. and Sarro, Federica and Harman, Mark},
    month = apr,
    year = {2024},
    note = {Conference Name: 2024 IEEE/ACM 46th International Conference on Software Engineering (ICSE)
Meeting Name: 2024 IEEE/ACM 46th International Conference on Software Engineering (ICSE)
Place: Lisbon, Portugal
Publisher: IEEE/ACM
Volume: 46},
}

@article{li_bias_2025,
    title = {Bias behind the wheel: {Fairness} testing of autonomous driving systems},
    volume = {34},
    number = {3},
    journal = {ACM Transactions on Software Engineering and Methodology},
    author = {Li, Xinyue and Chen, Zhenpeng and Zhang, Jie M and Sarro, Federica and Zhang, Ying and Liu, Xuanzhe},
    year = {2025},
    note = {Publisher: ACM New York, NY},
    pages = {1--24},
}

@article{canali_challenges_2022,
    title = {Challenges and recommendations for wearable devices in digital health: {Data} quality, interoperability, health equity, fairness},
    volume = {1},
    issn = {2767-3170},
    shorttitle = {Challenges and recommendations for wearable devices in digital health},
    url = {https://journals.plos.org/digitalhealth/article?id=10.1371/journal.pdig.0000104},
    doi = {10.1371/journal.pdig.0000104},
    language = {en},
    number = {10},
    urldate = {2025-09-14},
    journal = {PLOS Digital Health},
    author = {Canali, Stefano and Schiaffonati, Viola and Aliverti, Andrea},
    year = {2022},
    note = {Publisher: Public Library of Science},
    pages = {e0000104},
}

@article{daloisio_how_2024,
  title={How fair are we? From conceptualization to automated assessment of fairness definitions},
  author={d’Aloisio, Giordano and Di Sipio, Claudio and Di Marco, Antinisca and Di Ruscio, Davide},
  journal={Software and Systems Modeling},
  pages={1--27},
  year={2025},
  publisher={Springer Berlin Heidelberg}
}

@article{austin_will_2016,
    title = {Will {I} {Pass} the {Bar} {Exam}: {Predicting} {Student} {Success} {Using} {LSAT} {Scores} and {Law} {School} {Performance}},
    volume = {45},
    journal = {HofstrA l. rev.},
    author = {Austin, Katherine A and Christopher, Catherine Martin and Dickerson, Darby},
    year = {2016},
    note = {Publisher: HeinOnline},
    pages = {753},
}

@inproceedings{sarro2023search,
  title={Search-based software engineering in the era of modern software systems},
  author={Sarro, Federica},
  booktitle={Proceedings of the IEEE International Conference on Requirements Engineering},
  volume={2023},
  pages={3--5},
  year={2023},
  organization={Institute of Electrical and Electronics Engineers (IEEE)}
}

@inproceedings{hort2021fairea,
  title={Fairea: A model behaviour mutation approach to benchmarking bias mitigation methods},
  author={Hort, Max and Zhang, Jie M and Sarro, Federica and Harman, Mark},
  booktitle={Proceedings of the 29th ACM joint meeting on European software engineering conference and symposium on the foundations of software engineering},
  pages={994--1006},
  year={2021}
}

@article{chen2023comprehensive,
  title={A comprehensive empirical study of bias mitigation methods for machine learning classifiers},
  author={Chen, Zhenpeng and Zhang, Jie M and Sarro, Federica and Harman, Mark},
  journal={ACM transactions on software engineering and methodology},
  volume={32},
  number={4},
  pages={1--30},
  year={2023},
  publisher={ACM New York, NY, USA}
}

@article{chen2024fairness,
  title={Fairness testing: A comprehensive survey and analysis of trends},
  author={Chen, Zhenpeng and Zhang, Jie M and Hort, Max and Harman, Mark and Sarro, Federica},
  journal={ACM Transactions on Software Engineering and Methodology},
  volume={33},
  number={5},
  pages={1--59},
  year={2024},
  publisher={ACM New York, NY}
}

@inproceedings{daloisio_democratizing_2023,
    address = {Cham},
    series = {Lecture {Notes} in {Computer} {Science}},
    title = {Democratizing {Quality}-{Based} {Machine} {Learning} {Development} through {Extended} {Feature} {Models}},
    copyright = {All rights reserved},
    isbn = {978-3-031-30826-0},
    doi = {10.1007/978-3-031-30826-0_5},
    language = {en},
    booktitle = {Fundamental {Approaches} to {Software} {Engineering}},
    publisher = {Springer Nature Switzerland},
    author = {d’Aloisio, Giordano and Di Marco, Antinisca and Stilo, Giovanni},
    editor = {Lambers, Leen and Uchitel, Sebastián},
    year = {2023},
    pages = {88--110},
}

@article{hort_multi-objective_2023,
    title = {Multi-objective search for gender-fair and semantically correct word embeddings},
    volume = {133},
    issn = {1568-4946},
    url = {https://www.sciencedirect.com/science/article/pii/S1568494622009656},
    doi = {10.1016/j.asoc.2022.109916},
    urldate = {2025-01-23},
    journal = {Applied Soft Computing},
    author = {Hort, Max and Moussa, Rebecca and Sarro, Federica},
    month = jan,
    year = {2023},
    pages = {109916},
}

@article{perera_search-based_2022,
    title = {Search-based fairness testing for regression-based machine learning systems},
    volume = {27},
    issn = {1573-7616},
    url = {https://doi.org/10.1007/s10664-022-10116-7},
    doi = {10.1007/s10664-022-10116-7},
    language = {en},
    number = {3},
    urldate = {2023-04-04},
    journal = {Empirical Software Engineering},
    author = {Perera, Anjana and Aleti, Aldeida and Tantithamthavorn, Chakkrit and Jiarpakdee, Jirayus and Turhan, Burak and Kuhn, Lisa and Walker, Katie},
    month = mar,
    year = {2022},
    pages = {79},
}

@article{kozodoi_fairness_2022,
    title = {Fairness in credit scoring: {Assessment}, implementation and profit implications},
    volume = {297},
    issn = {0377-2217},
    doi = {10.1016/J.EJOR.2021.06.023},
    number = {3},
    journal = {European Journal of Operational Research},
    author = {Kozodoi, Nikita and Jacob, Johannes and Lessmann, Stefan},
    month = mar,
    year = {2022},
    note = {Publisher: North-Holland},
    pages = {1083--1094},
}

@misc{noauthor_eu_2023,
    title = {{EU} {AI} {Act}: first regulation on artificial intelligence {\textbar} {News} {\textbar} {European} {Parliament}},
    shorttitle = {{EU} {AI} {Act}},
    url = {https://www.europarl.europa.eu/news/en/headlines/society/20230601STO93804/eu-ai-act-first-regulation-on-artificial-intelligence},
    language = {en},
    urldate = {2024-01-05},
    month = aug,
    year = {2023},
}

@inproceedings{gong_greenstableyolo_2024,
    title = {{GreenStableYolo}: {Optimizing} {Inference} {Time} and {Image} {Quality} of {Text}-to-{Image} {Generation}},
    booktitle = {International {Symposium} on {Search} {Based} {Software} {Engineering}},
    publisher = {Springer Nature Switzerland Cham},
    author = {Gong, Jingzhi and Li, Sisi and d’Aloisio, Giordano and Ding, Zishuo and Ye, Yulong and Langdon, William B and Sarro, Federica},
    year = {2024},
    pages = {70--76},
}

@inproceedings{d2025manila,
  title={MANILA: A Low-Code Application to Benchmark Machine Learning Models and Fairness-Enhancing Methods},
  author={d'Aloisio, Giordano},
  booktitle={Proceedings of the 33rd ACM International Conference on the Foundations of Software Engineering},
  pages={1153--1157},
  year={2025}
}

@article{d2025towards,
  title={Towards early detection of algorithmic bias from dataset’s bias symptoms: An empirical study},
  author={d’Aloisio, Giordano and Di Sipio, Claudio and Di Marco, Antinisca and Di Ruscio, Davide},
  journal={Information and Software Technology},
  pages={107905},
  year={2025},
  publisher={Elsevier}
}

\end{document}